\documentclass{article}
\usepackage[english]{babel}

\usepackage[a4paper,top=2cm,bottom=2cm,left=3cm,right=3cm,marginparwidth=1.75cm]{geometry}

\usepackage{amsmath}
\usepackage{graphicx}
\usepackage[colorlinks=true, allcolors=blue]{hyperref}
\usepackage{booktabs}
\usepackage{pifont}
\usepackage{array}
\usepackage{siunitx}
\usepackage{float}
\floatstyle{ruled}
\newfloat{algorithm}{tbp}{loa}
\floatname{algorithm}{Algorithm}

\usepackage[table]{xcolor}
\usepackage{amsfonts}
\usepackage{multirow}

\usepackage{authblk}

\usepackage{tikz}
\usetikzlibrary{shapes.geometric, arrows.meta, positioning}

\newcommand{\maybeincludegraphics}[2][]{%
  \IfFileExists{#2}{%
    \includegraphics[#1]{#2}%
  }{%
    \fbox{\parbox[c][0.18\textheight][c]{0.95\linewidth}{\centering Missing figure:\\\texttt{\detokenize{#2}}}}%
  }%
}

\tikzstyle{startstop} = [rectangle, rounded corners, minimum width=3.5cm, minimum height=1cm,text centered, draw=black, fill=blue!20]
\tikzstyle{process} = [rectangle, minimum width=3.5cm, minimum height=1cm, text centered, draw=black, fill=orange!20]
\tikzstyle{cluster} = [rectangle, minimum width=3.5cm, minimum height=1cm, text centered, draw=black, fill=green!20]
\tikzstyle{arrow} = [thick,->,>=stealth]

\usepackage{cite}
\usepackage{amssymb,amsfonts}
\usepackage{algorithmic}
\usepackage{textcomp}
\usepackage{xcolor}
\def\BibTeX{{\rm B\kern-.05em{\sc i\kern-.025em b}\kern-.08em
    T\kern-.1667em\lower.7ex\hbox{E}\kern-.125emX}}

\title{Nf-PEAK: Process-Based Energy Attribution for Nextflow Workflows on Kubernetes Clusters%
}

\author[1]{Philipp Thamm}
\author[2]{Somayeh Mohammadi}
\author[3]{Kathleen West}
\author[2]{Knut Reinert}
\author[3]{Lauritz Thamsen}
\author[1]{Ulf Leser}

\affil[1]{Humboldt-Universität zu Berlin, Berlin, Germany\\
\texttt{\{thammphx,ulf.leser\}@hu-berlin.de}}

\affil[2]{Freie Universität Berlin, Berlin, Germany\\ \newline
\texttt{mohamadi.s@gmail.com}, \texttt{knut.reinert@fu-berlin.de}}

\affil[3]{University of Glasgow, Glasgow, United Kingdom\\
\texttt{\{Kathleen.West,lauritz.thamsen\}@glasgow.ac.uk}}

\date{} % <-- removes current date

\begin{document}

\maketitle

\begin{abstract}
Scientific workflows are pipelines of interdependent tasks. 
They are increasingly executed on shared Kubernetes clusters via workflow engines such as Nextflow.
Their energy consumption matters for both cost and sustainability.
It is necessary to examine and optimize workflow tasks individually, because they can be very heterogeneous. 
However, estimating task-level energy on clusters is difficult: Intel RAPL counters report only node-level energy, access to counters and host process information is typically restricted, and concurrent workloads introduce resource contention and measurement noise.

We present \textbf{Nf-PEAK}, a containerized method to attribute CPU-package and DRAM energy to individual \textit{processes} and Nextflow \textit{tasks}.
Nf-PEAK (i) identifies workflow pods, (ii) maps pods to host processes via cgroup metadata, (iii) samples RAPL and per-process performance counters, and (iv) applies a non-linear energy-credit model before aggregating results at task level.
On a Kubernetes cluster, we evaluate three nf-core workflows under controlled co-located CPU load.
Nf-PEAK reaches an average Mean Absolute Percentage Error of 6.6\% in isolated runs and 10.9\% when an unrelated workload saturates 8 of 32 hardware threads per node, and remains stable across 2, 3, 4, and 8 nodes.
Compared to the state-of-the-art Kubernetes tool Kepler, Nf-PEAK yields lower error on average, particularly under co-located load.
\end{abstract}

%\begin{IEEEkeywords}
%scientific workflows, infrastructure profiling, energy measurement, cluster computing, cloud computing, sustainable computing
%\end{IEEEkeywords}

\section{Introduction}\label{Introduction}
Extracting information from a dataset can require multiple interdependent processing steps.
Such pipelines of tasks are referred to as \emph{scientific workflows}~\cite{dayal_scientific_2009} and are used in many areas, such as genomics~\cite{fellows_yates_reproducible_2021} or remote sensing~\cite{sudmanns_assessing_2020}.
Workflow configurations are commonly optimized for fast execution, but recently energy-efficient computing has received more attention due to environmental concerns and rising energy costs~\cite{freitag_real_2021, de_roucy-rochegonde_ai_2025, batz_lineiro_pay-back_2025}.
For example, one run of the workflow Sarek\footnote{https://nf-co.re/sarek/3.5.1/, last accessed: March 22, 2026} for genome variant detection consumes more than 1.16\,MJ of energy across 4 cluster nodes over a runtime of 56 minutes in our experiments with 10.6\,GB of real-world input data. Of this energy, about 0.7\,MJ is attributable to the workflow itself after subtracting static energy consumption of the cluster and overhead introduced by the Kubernetes scheduler, equivalent to drawing an average power of 208.3\,W over the workflow's runtime.

Optimizing a workflow for energy efficiency requires feedback beyond node-level totals: workflow developers and users need to understand \emph{which tasks} dominate energy consumption and how configuration changes affect them.
A common approach to obtain energy information on Intel systems is to use Running Average Power Limit (RAPL) energy counters~\cite{david_rapl_2010, khan_rapl_2018}, which provide low-overhead energy readings for CPU and (depending on the CPU model) DRAM.
While access to RAPL is straightforward on local hardware through tools such as perf\footnote{https://perfwiki.github.io/main/, last accessed: March 22, 2026} or powercap\footnote{https://docs.kernel.org/power/powercap/powercap.html, last accessed: March 22, 2026}, extracting accurate \emph{task-level} information on managed clusters is challenging for several reasons:

\begin{itemize}
    \item \textbf{Limited access:} User privileges are often limited, making it impossible to install supporting tools on nodes or to access RAPL counters and host-wide process information.
    \item \textbf{Containerization:} Cluster workloads are commonly containerized, which restricts access to RAPL and to processes running outside the current container. %unless special privileges are granted.
    \item \textbf{Resource manager:} Workloads are scheduled on the cluster by a resource manager sitting between workflow engine and nodes, preventing knowledge about the location of individual tasks prior to execution.
    \item \textbf{Distributed execution:} Workflows are executed across multiple nodes, so measurement must be coordinated and aggregated.
    \item \textbf{Complex node architecture:} Multi-socket CPUs and NUMA memory require socket-aware monitoring and attribution.
    \item \textbf{Shared environment:} Multiple workloads may run on the same nodes concurrently, but RAPL only counts at CPU package level, aggregating energy across unrelated workloads.
\end{itemize}

%As a result, monitoring total node energy provides limited insights for heterogeneous workflows, whose tasks can vary widely in runtime and resource usage~\cite{fellows_yates_reproducible_2021}.
%Instead, a method that attributes measured node energy to individual \emph{workflow tasks} is required to enable energy-aware workflow optimization.

In this paper, we present \textbf{Nf-PEAK} (\textbf{P}rocess-based \textbf{E}nergy \textbf{A}ttribution on \textbf{K}ubernetes clusters targeting \textbf{N}ext\textbf{f}low workflows), a method to estimate the energy consumption of individual processes and workflow tasks on Kubernetes clusters.
In summary, Nf-PEAK combines RAPL measurements with per-process performance counters and uses a non-linear attribution strategy inspired by EnergAt~\cite{he_energat_2024}.
The tool is fully containerized and can be deployed through Kubernetes without the need to install special software on cluster nodes.

Our contributions are:
\begin{itemize}
    \item \textbf{Containerized task attribution on Kubernetes:} a practical pipeline to discover Nextflow tasks, map them to host processes, and monitor CPU and DRAM energy on all involved nodes.
    \item \textbf{Process-level non-linear attribution:} an energy-credit model adapted to process granularity, enabling accurate CPU and DRAM energy attribution for heterogeneous tasks and concurrent cluster workloads.
    \item \textbf{Evaluation on real workflows:} experiments with three nf-core workflows across four cluster sizes (2, 3, 4, and 8 nodes) and under controlled co-located CPU load, including a comparison to Kepler~\cite{amaral_kepler_2023, amaral_process-based_2024}.
\end{itemize}

In the following, we first provide background about the technologies used in our work. %in Section \ref{Background}. 
In Section \ref{Attribution_Strategy}, we present how Nf-PEAK attributes energy consumption and integrates with Kubernetes. Our experiments and their results are presented in Section \ref{Experiments} and discussed in Section \ref{Discussion}. Section \ref{RelatedWork} presents related work in the area of software-based energy attribution. %Open challenges for future work are explained in Section \ref{Future_Work}. 
Finally, Section \ref{Conclusion} concludes the paper.

\section{Background}\label{Background}
This section summarizes the technologies and concepts most relevant to our work: Executing scientific workflows on Kubernetes, RAPL energy counters, and energy attribution strategies.

\subsection{Scientific Workflows on Kubernetes}
\textbf{Kubernetes}~\cite{carrion_kubernetes_2023} is a widely used container orchestration system that coordinates the execution of workloads, including scientific workflows, on compute clusters. It implements scheduling and re-scheduling of tasks on a distributed infrastructure, and also supports automatic resource allocation and load balancing.

Workloads are executed on the nodes through containers, packaging programs together with their dependencies and execution environment. These containers are running in isolation on top of the node's operating system's kernel. The containers utilized by Kubernetes are commonly built and managed by Docker~\cite{rad_introduction_2017}. They are organized in Kubernetes pods. Each pod contains one or multiple containers and is scheduled independently.

\textbf{Scientific workflows} decompose an analysis into separate logical tasks with explicit data dependencies~\cite{gil_examining_2007}.
Nextflow~\cite{di_tommaso_nextflow_2017} expresses logical tasks in a DSL and orchestrates their execution in the form of physical tasks on a variety of backends, including Kubernetes. %\cite{carrion_kubernetes_2023}.
Here, each logical task is split into one or multiple physical tasks, depending on the task and workflow configuration.
In a Kubernetes backend, Nextflow executes each physical workflow task in an isolated pod/container (e.g., Docker%~\cite{rad_introduction_2017}
), using Kubernetes to orchestrate execution on a compute cluster~\cite{van_steen_brief_2016} and to manage the available resources.
%In this paper, we attribute energy to \emph{processes} running on the nodes and aggregate results to \emph{workflow tasks}.

\subsection{RAPL Energy Counters}
Intel RAPL exposes energy counters for different domains (depending on CPU model), such as \emph{Package} (CPU package) and \emph{DRAM}~\cite{khan_rapl_2018}.
RAPL readings are low-overhead and widely used for power/energy estimation~\cite{hackenberg_energy_2015}.
However, RAPL counters are \emph{coarse-grained}: each CPU provides only aggregate energy for all workloads on that CPU and directly associated memory.
In shared clusters, this makes it impossible to directly infer task-level energy from RAPL alone.
Furthermore, energy counters are bounded registers and overflow after a certain amount of counted energy, requiring periodic sampling to compute correct deltas over longer durations. On our hardware, the energy counted before an overflow is 262,143.3\,J for CPUs and 65,713\,J for DRAM.

\subsection{Energy Attribution and EnergAt}
Energy attribution methods allocate measured energy to specific workloads, at the level of containers, processes, or threads, based on resource usage.
A common distinction is between \emph{static} energy (idle baseline) and \emph{dynamic} energy induced by workload execution.
While dynamic energy can be attributed based on activity, attributing static energy depends on the usage scenario: according to the guidelines stipulated in the Greenhouse Gas Protocol\footnote{https://ghgprotocol.org/sites/default/files/2023-03/GHGP-ICTSG\%20-\%20ALL\%20Chapters.pdf, last accessed: March 22, 2026}, static energy should be apportioned among concurrent workloads proportionally to their resource usage in shared clusters.

EnergAt~\cite{he_energat_2024} is a NUMA-aware energy attribution method that uses \emph{energy credits} derived from CPU-time fractions and a non-linear scaling factor to model the power--utilization relationship.
Nf-PEAK adapts this idea to a Kubernetes setting and to \emph{process-level} monitoring, enabling task-level aggregation for workflow engines.

%\section{Attribution Strategy}\label{Attribution_Strategy}
\section{Method}\label{Attribution_Strategy}
This section describes Nf-PEAK, our approach for task-level energy attribution for scientific workflows executed on compute clusters.
\subsection{Approach Overview}\label{Approach_Overview}
\begin{figure}
\centering
\maybeincludegraphics[width=1.00\linewidth, trim={2.0cm 1.0cm 1.0cm 2.0cm},clip]{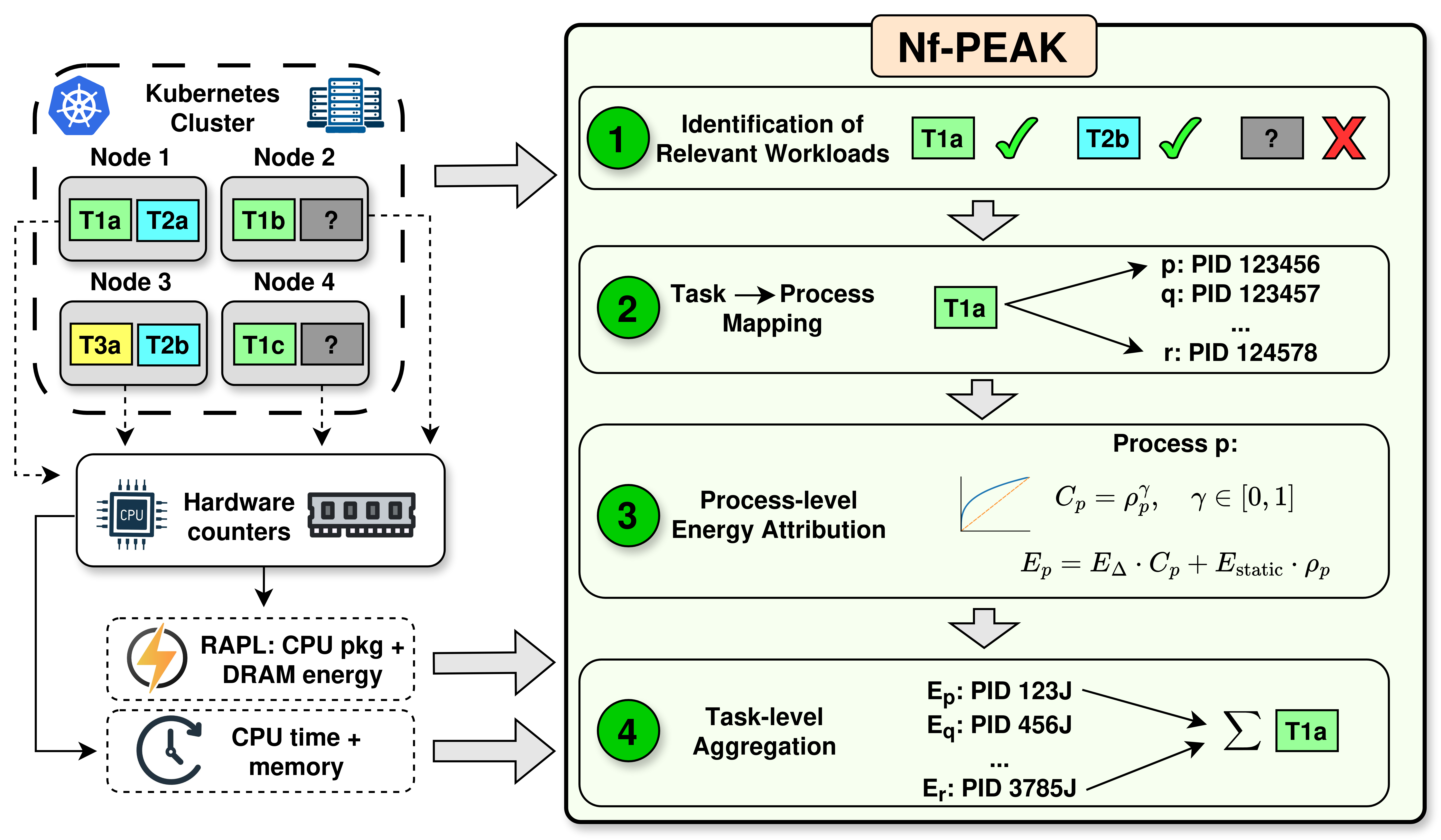}
\caption{Overview of energy attribution with Nf-PEAK. %High-level design of Nf-PEAK with per-task energy attribution based on a non-linear metric computing per-process estimations. Nf-PEAK uses 
%Using three types of input data presented on the left, including information about currently running pods from Kubernetes, node-level energy data and process-level CPU time and memory statistics, Nf-PEAK attributes task-level energy in four steps. %The output consists of attributions on task-level, including physical and logical tasks, as well as additional workflow-level and process-level energy data.
Using pod information from Kubernetes, node-level energy data and process-level CPU time and memory statistics, Nf-PEAK attributes task-level energy in four steps.
}
\label{fig:Method}
\end{figure}

Scientific workflows commonly contain multiple tasks that can be executed in parallel. Additionally, unrelated workloads are often run simultaneously on the same hardware in shared %Kubernetes 
clusters.
To determine the energy consumption of individual workflow tasks, Nf-PEAK performs energy attribution.
%Nf-PEAK is implemented as a containerized pipeline with two complementary viewpoints: (a) a user-side controller that can observe pods cluster-wide via the Kubernetes API, and (b) node-local monitoring pods that can read RAPL counters and inspect host process metadata.
An overview of Nf-PEAK is presented in Figure \ref{fig:Method}.
Using three types of input data, including information about currently running pods, node-level energy data and process-level CPU time and memory statistics, Nf-PEAK attributes task-level energy in four steps.
During a workflow run, Nf-PEAK (i) identifies %Nextflow 
task pods and extracts metadata (task name, pod UID), (ii) maps each pod UID to the corresponding host PIDs via %cgroup 
process metadata, (iii) %samples RAPL (CPU Package, DRAM) and per-process counters from \texttt{/proc} at fixed intervals, and (iv) allocates each interval's energy to concurrent processes using a static/dynamic split and a non-linear credit model before summing per-process estimates to per-task and per-workflow results. %An overview of Nf-PEAK is presented in Figure \ref{fig:Method}.
attributes per-process energy using a non-linear credit model, %based on process metadata and node-level energy counters, 
and (iv) aggregates process-level energy to produce task-level results.

\subsection{Energy Attribution}\label{Attribution}
%We compute task-level energy attributions by calculating the sum of the energy attributed to each of the processes executed as part of a workflow task. 
We compute energy attributions for each individual workflow task by calculating the sum of the energy attributed to each of its processes.
To attribute energy to an individual process, we use the following approach:
\textbf{Static and dynamic energy.} Let \(E^{D}_{\text{total},s}\) be the total energy in domain \(D\in\{\text{CPU,DRAM}\}\) measured by RAPL on socket \(s\) during a sampling interval of length \(T\).
Nf-PEAK estimates static (idle) power \(P^{D}_{\text{static},s}\) by sampling RAPL at some point before workflow execution while the node is idle and computes static energy during an interval as:
%\[
\begin{equation}
E^{D}_{\text{static},s} = P^{D}_{\text{static},s}\cdot T
\end{equation}
%\]
Dynamic energy is then:
%\[
\begin{equation}
E^{D}_{\Delta,s}=E^{D}_{\text{total},s}-E^{D}_{\text{static},s}
\end{equation}
%\]

%\subsubsection*{Energy credits and non-linearity}
\textbf{CPU energy.} For a process \(a\), we estimate its CPU-time share on socket \(s\) within the interval as
\( \rho_{a,s} = T^{\text{CPU}}_{a,s} / T^{\text{CPU}}_{\text{total},s}\),
where \(T^{\text{CPU}}_{\text{total},s}\) is the total CPU time accrued by all processes on \(s\).
Note that unlike EnergAt~\cite{he_energat_2024}, we are directly monitoring processes instead of individual hardware threads and compute energy credits based on CPU time share of processes.
We compute an \emph{energy credit} as:
%\[
\begin{equation}
C^{\text{CPU}}_{a,s} = \rho_{a,s}^{\gamma}, \quad \gamma\in[0,1]
\end{equation}
%\]
where \(\gamma\) is a hyperparameter modeling the non-linear power--utilization relationship (e.g., due to DVFS and other hardware effects).
CPU energy is attributed as:
%\[
\begin{equation}
\label{eq4}
E^{\text{CPU}}_{a} = \sum_{s\in S}\left(E^{\text{CPU}}_{\Delta,s}\cdot C^{\text{CPU}}_{a,s} + E^{\text{CPU}}_{\text{static},s}\cdot \rho_{a,s}\right)
\end{equation}
%\]
%For DRAM energy attribution, NUMA locality is additionally taken into account, following the design of EnergAt~\cite{he_energat_2024}.

\textbf{Memory energy.} To attribute energy to a process based on its memory usage, we use the same approach as for the energy used by the CPU, but leveraging the DRAM domain of RAPL and the fraction of the total memory belonging to the process instead. For a process \(a\), we estimate its memory share on socket \(s\) within the interval as
\( \sigma_{a,s} = M_{a,s} / M_{\text{total},s}\),
where \(M_{\text{total},s}\) is the total memory space accrued by all processes on \(s\). An energy credit is then again computed as:
%\[
\begin{equation}
C^{\text{M}}_{a,s} = \sigma_{a,s}^{\gamma}, \quad \gamma\in[0,1]
\end{equation}
%\]
similar to the credit for CPU time. Memory energy is attributed as:
%\[
\begin{equation}
\label{eq6}
E^{\text{M}}_{a} = \sum_{s\in S}\left(E^{\text{M}}_{\Delta,s}\cdot C^{\text{M}}_{a,s} + E^{\text{M}}_{\text{static},s}\cdot \sigma_{a,s}\right)
\end{equation}
%\]
\textbf{Total process energy.} The final energy attributed to a process is the sum of the energy attributed to CPU package and memory:
%\[
\begin{equation}
E_{a} = E^{\text{CPU}}_{a} + E^{\text{M}}_{a}
\end{equation}

\subsection{Hyperparameters}
Nf-PEAK has three main parameters: the non-linearity exponent \(\gamma\), the polling interval for detecting new tasks/processes, and the RAPL sampling interval.

\begin{figure}[t]
\centering
\maybeincludegraphics[width=0.68\linewidth, trim={0.0cm 0.0cm 0.0cm 0.7cm},clip]{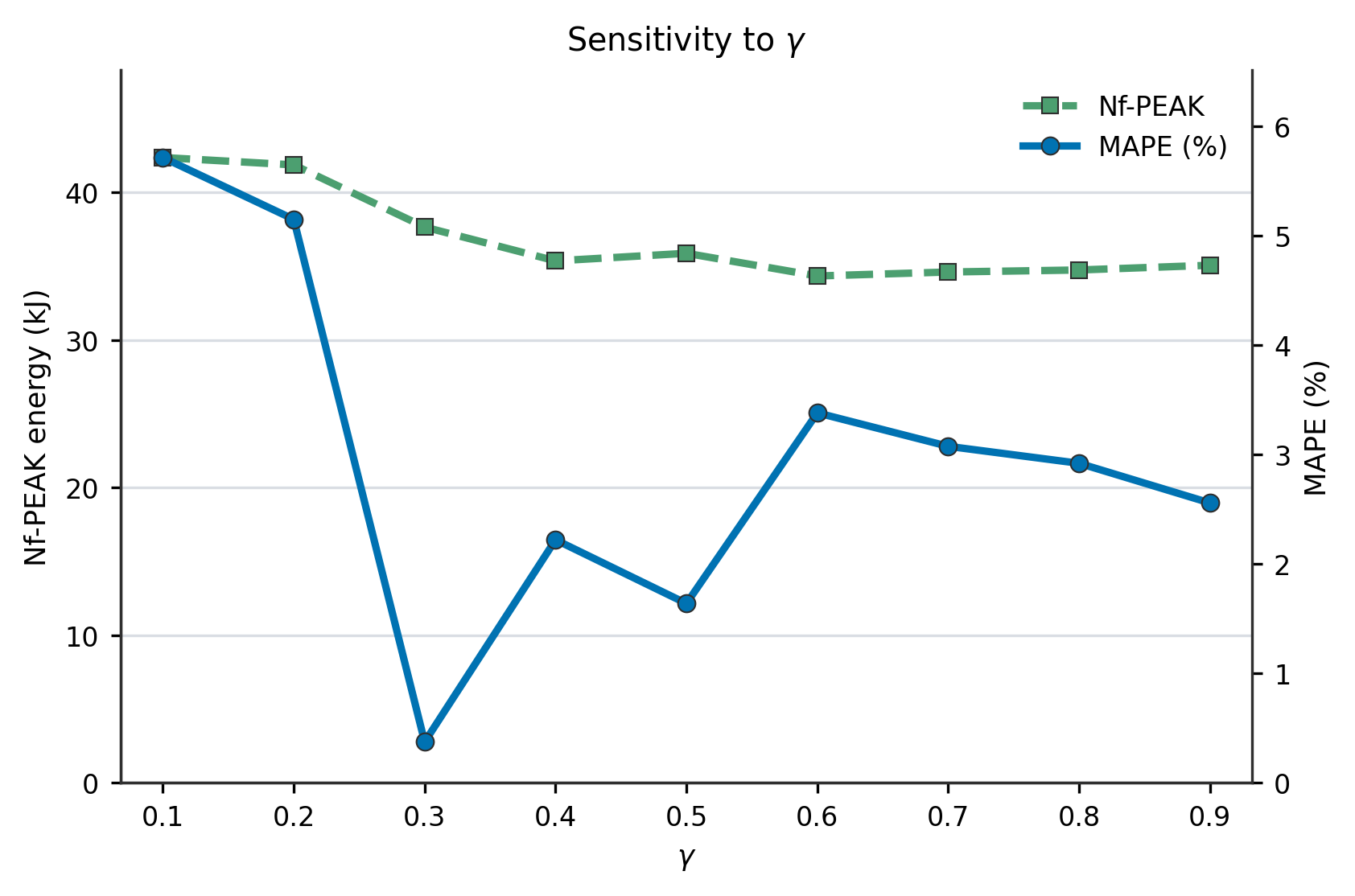}
\caption{Sensitivity of attributed Nf-PEAK energy and MAPE to the non-linearity exponent \(\gamma\) on the synthetic calibration workflow.
The minimum MAPE occurs at \(\gamma=0.3\), which is used for all experiments.}
\label{fig:gamma_sensitivity}
\end{figure}

The non-linearity exponent \(\gamma\) controls how strongly resource shares are transformed into energy credits and thereby captures the non-linear relationship between power and CPU time/memory utilization. In our experiments, we set \(\gamma=0.3\). This value was calibrated on the target CPUs by running a representative synthetic workflow and attributing energy using Nf-PEAK with \(\gamma\in\{0.1,0.2,0.3,0.4,0.5,0.6,0.7,0.8,0.9\}\). The workflow contains a series of tasks executed sequentially that use \texttt{stress-ng} to introduce varying load on the CPU, ranging from completely idle to full load on all cores in steps of 10\% CPU load. We evaluated attribution quality using the MAPE defined in Section~\ref{ExperimentalDesign}, which was lowest for \(\gamma=0.3\). To avoid overfitting \(\gamma\) to the synthetic workflow used for calibration, we did not conduct additional fine-tuning experiments with values around \(\gamma=0.3\) to further optimize \(\gamma\) on this workflow. The synthetic workflow used for calibration is not part of our experimental evaluation. Figure~\ref{fig:gamma_sensitivity} shows the calibration experiment used to select \(\gamma=0.3\).

The polling interval determines how often Nf-PEAK checks for newly started tasks/processes. In our experiments, we poll for new tasks/processes every 5\,s. Section~\ref{ExperimentalResults} includes an experiment that quantifies the trade-off between shorter polling intervals and increased monitoring overhead.

The RAPL sampling interval determines the temporal granularity at which Nf-PEAK reads the RAPL energy counters. In our experiments, we sample RAPL every 2\,s. This RAPL sampling interval was evaluated in the same experiment as the polling interval.

\subsection{Architecture and Deployment}\label{Architecture}
\begin{figure}
%\centering
\maybeincludegraphics[width=1.00\linewidth, trim={0.0cm 90cm 0.0cm 0.0cm},clip]{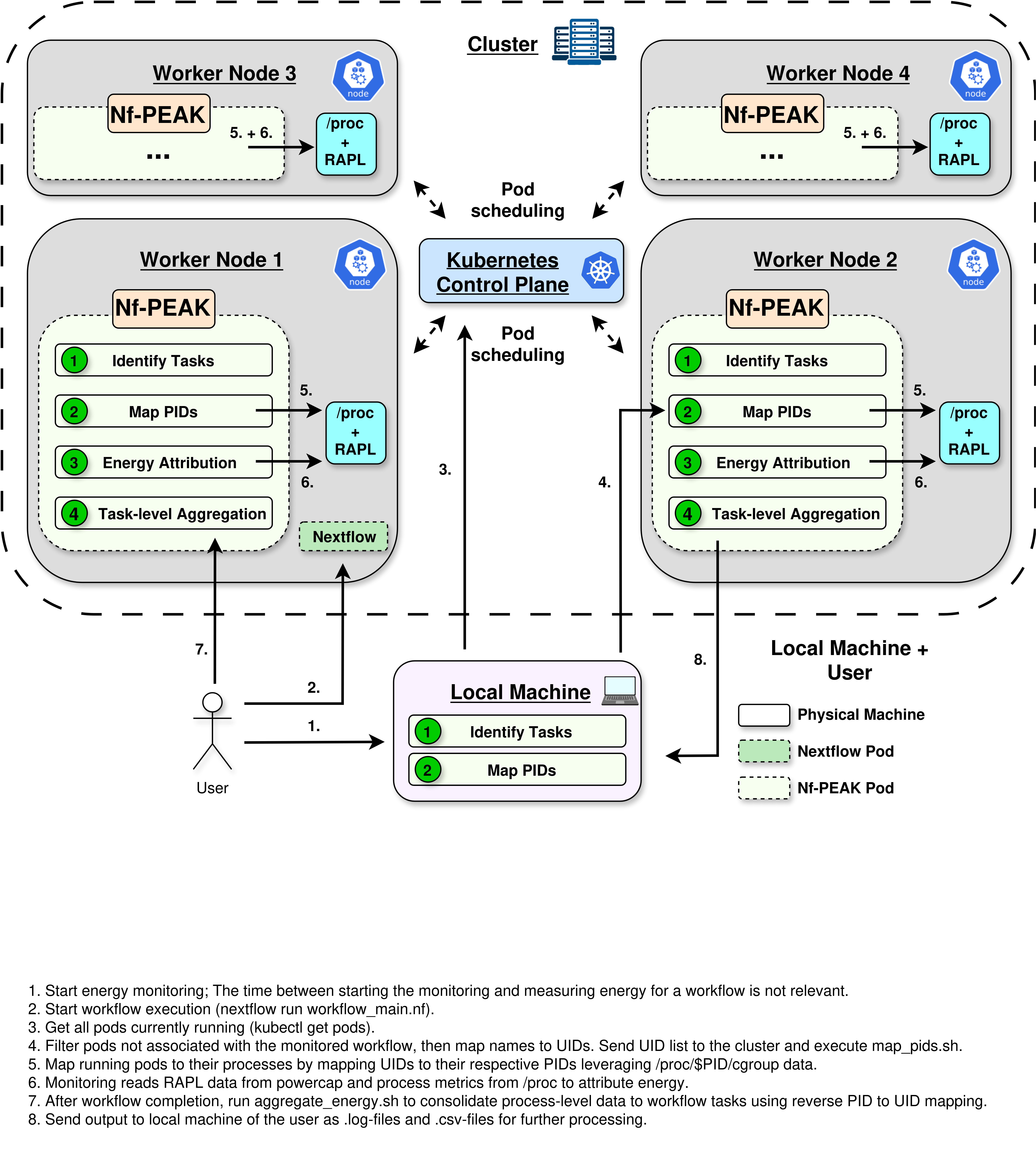}
\caption{Actors and components of Nf-PEAK and their communication during attribution. Steps are numbered chronologically. Energy monitoring may start any time before workflow execution. %(the time between starting monitoring and measuring a workflow is irrelevant). 
Numbered steps include the following actions:
1. User starts energy monitoring.
2. User starts workflow execution.
3. Query currently running pods.
4. Filter associated pod UIDs and send the UID list to the cluster.
5. Map pods to associated processes using \mbox{\texttt{/cgroup}}.
6. Attribute energy for newly identified PIDs by reading RAPL counters from \texttt{powercap} and process metrics from \texttt{/proc}. %to attribute energy.
7. %After workflow completion, c
Consolidate process-level data to task-level using PID$\rightarrow$UID mapping.
8. Transfer results %the resulting \texttt{.log} and \texttt{.csv} files 
for post-processing.}
\label{fig:Architecture}
\end{figure}

Nf-PEAK is deployed as a set of containers through Kubernetes.
%Some steps (e.g., starting monitoring and collecting results) can be triggered from a user machine, while monitoring and mapping must be performed on the cluster nodes.
The deployment separates a user-side controller from node-local monitoring components.
The controller starts and stops monitoring, queries the Kubernetes API for running workflow pods, filters pods that belong to the monitored Nextflow execution, and collects the final result files.
The node-local monitoring components run on the cluster nodes where the workflow pods are executed.
They map pod UIDs to host PIDs, sample RAPL counters and process metrics, and write process-level attribution data for later aggregation.
Figure~\ref{fig:Architecture} summarizes the actors and the data flow of Nf-PEAK.

During execution, the controller periodically discovers newly started task pods and forwards their UIDs to the monitoring components on the nodes where the pods are scheduled.
Each monitoring component inspects cgroup metadata on its respective node in \texttt{/proc} to identify the host PIDs associated with the received pod UIDs.
For each discovered PID, per-process CPU time and memory statistics are sampled and combined with RAPL Package and DRAM energy deltas from the same node.
After the workflow finishes, the collected PID-level records are mapped to the UID of their physical workflow tasks and aggregated to physical tasks, logical tasks, and workflow-level summaries.

Only the monitoring pods require visibility of all PIDs on the host, realized with \texttt{hostPID: true}.
Workflow pods remain unmodified and do not require additional privileges, instrumentation, or changes to the Nextflow workflow definition.
This separation keeps the monitored workflow portable while limiting access to host-level information to a small set of known monitoring components.
The security implications of this deployment model are discussed in Section~\ref{Discussion}.

\section{Evaluation}\label{Experiments}
We evaluate (i) attribution accuracy in isolated executions, (ii) stability across cluster sizes, (iii) robustness under co-located CPU load, and (iv) performance relative to Kepler.
%~\cite{amaral_kepler_2023}.

\subsection{Hardware and Software Setup}\label{HardwareSetup}
We use a Kubernetes cluster consisting of 26 nodes in total, of which 8 were available exclusively for our experiments.
Eight nodes are sufficient to provide distributed environments for experiments under conditions equivalent to a real-world cluster environment.
At the same time, using more than 8 nodes would reduce node-level task parallelism in the tested workflows.
Since providing accurate energy attribution for parallel workflow tasks is the primary goal of our work, creating conditions where parallel tasks occur regularly is important for a thorough evaluation.
Each node contains an Intel(R) Xeon(R) Silver 4314 CPU (2.40\,GHz base, 3.40\,GHz max) and supports the RAPL Package and DRAM domains.
The cluster uses Kubernetes for orchestration and deploys Docker containers.
Workflows are implemented and executed in Nextflow.
%~\cite{di_tommaso_nextflow_2017}.

\subsection{Workflows}\label{UsedWorkflows}
We evaluate three nf-core workflows~\cite{philip_a_ewels_nf-core_2020} 
%\footnote{https://nf-co.re/pipelines/, last accessed: March 22, 2026} 
that cover different task granularities and parallelism patterns.
Table~\ref{tab:Workflow_Table} provides a qualitative overview of the workflows and summarizes key execution metrics in our experimental configuration.%, and Table~\ref{tab:Wf_Overview_Table} provides a qualitative overview.

%\begin{table}[h]
%    \centering
%    \caption{Qualitative overview of workflow characteristics (configuration- and data-dependent).}
%    \label{tab:Wf_Overview_Table}
%    \begin{tabular}{lccccc}
%        \toprule
%        \textbf{Workflow} & \textbf{Domain} & \textbf{Goal} & \textbf{Parallelism} & \textbf{Task Len.} & \textbf{CPU usage}\\
%        \midrule
%        RNASeq      & Bioinf. & RNA sequencing & Low & Long & Mostly low\\
%        Sarek       & Bioinf. & Genome var. calling & Medium & Medium & High\\
%        Rangeland   & Rem. Sensing & Land-cover analysis & High & Short & High\\
%        \bottomrule
%    \end{tabular}
%\end{table}

\begin{table*}[h]
    \centering
    \caption{Key metrics of the workflows in our experiments for an execution on 2 nodes. CPU utilization metrics are given per task. %without considering parallel task execution.
    AvgCPU is the average CPU utilization over all individual tasks of the workflow, %not considering runtime, 
    while MaxCPU is the maximum average CPU utilization of any single task.
    %Executing parallel tasks on the same node can lead to higher total CPU utilization on that node.
    All memory and I/O metrics are totals over the whole workflow and across all used nodes.
    %The provided values are for an execution on two nodes.%, but they are very similar for more nodes since the number of nodes does not affect individual tasks directly. Only Makespan is reduced depending on the workflow.
    }
    \label{tab:Workflow_Table}
    \resizebox{\textwidth}{!}{%
    \begin{tabular}{lcccccccccc}
        \toprule
        \textbf{Workflow} & \textbf{Task Len.} & \textbf{Makespan} & \textbf{AvgCPU} & \textbf{MaxCPU} & \textbf{Memory} & \textbf{Read} & \textbf{Write} & \textbf{Tasks} & \textbf{DoP} \\
        \midrule
        RNASeq       & Long & 27\,min & 850\% & 1306\% & 168.2\,GB & 134.6\,GB & 85.3\,GB & 9 & 2\\
        Sarek       & Medium & 65\,min & 415\% & 2274\% & 1485.5\,GB & 454.4\,GB & 141.6\,GB & 152 & 24\\
        Rangeland   & Short & 48\,min & 190\% & 1246\% & 1340.2\,GB & 95.1\,GB & 21.3\,GB & 2665 & 100\\
        \bottomrule
    \end{tabular}
    }
\end{table*}

%\subsubsection*{RNASeq}
\textbf{RNASeq}\footnote{https://nf-co.re/rnaseq/3.18.0/, last accessed: March 22, 2026} is a bioinformatics workflow used to analyze RNA sequencing data with a reference genome.
RNASeq is characterized by few physical tasks and low task parallelism. %does not execute tasks concurrently on the same node.
%This enables a task-level comparison between individual RAPL measurements and Nf-PEAK attributions for non-overlapping tasks.

%\subsubsection*{Sarek}
\textbf{Sarek}\footnote{https://nf-co.re/sarek/3.5.1/, last accessed: March 22, 2026} is another bioinformatics workflow that detects variants on whole genome or targeted sequencing data.
Many tasks execute in parallel, causing high sustained CPU load.
%This makes Sarek suitable to evaluate attribution under high parallelism and high utilization.

%\subsubsection*{Rangeland}
\textbf{Rangeland}\footnote{https://nf-co.re/rangeland/1.0.0/, last accessed: March 22, 2026} is a remote sensing workflow that processes satellite imagery to assess land-cover changes.
It features numerous very short tasks, representing a worst case for the interval-based monitoring of Nf-PEAK.

\subsection{Experimental Design}\label{ExperimentalDesign}
We evaluate the accuracy of Nf-PEAK by comparing its task-level attributions
(aggregated to workflow level) against node-level RAPL measurements under controlled execution conditions.

For each workflow run, we collect in parallel:
(i) node-level energy by sampling the Package and DRAM RAPL counters on all involved nodes,
and (ii) task energy attributed by Nf-PEAK. %from per-process monitoring and subsequent task-level aggregation.
Because RAPL measures energy across all activity on a node, we compare primarily at the
workflow level using Mean Absolute Percentage Error (MAPE), detailed below.
To make the comparison meaningful, we account for (a) static energy not attributed by Nf-PEAK and
(b) platform/measurement overhead. %that is included in RAPL but not attributed to workflow tasks by Nf-PEAK.
%To assess how much the energy attributed by Nf-PEAK deviates from the expected values derived from RAPL, we compute Mean Absolute Percentage Error (MAPE), detailed below.

Unless stated otherwise, we repeat each experiment three times and report means. %Figures and tables report means across runs, except where individual runs are shown to visualize variance.

We conduct three sets of experiments that target complementary aspects of Nf-PEAK: 
(i) accuracy in an isolated setting, %where RAPL provides a strong reference signal, 
(ii) stability when increasing cluster size, and %changing the number of participating nodes, and 
(iii) robustness when unrelated workloads run concurrently.

\textbf{(1) Isolated runs.}
We execute RNASeq, Sarek, and Rangeland on 2 nodes without additional user workloads.
During each run, we measure node-level energy by sampling the Package and DRAM RAPL counters on all involved nodes and, in parallel, run Nf-PEAK to attribute energy to workflow tasks.
This isolated setup allows obtaining precise RAPL-based reference values for a direct comparison with Nf-PEAK.
% through MAPE.
%By subtracting the remaining static energy not attributed by Nf-PEAK and overhead from the RAPL measurement, this experiment allows a direct comparison of Nf-PEAK with a value based on a ground truth known to be accurate (RAPL).
%It is limited to scenarios with no co-located load, but it provides strong evidence about the accuracy of Nf-PEAK in this limited scenario.
By collecting RAPL measurements and Nf-PEAK attributions in parallel on the same runs, we ensure that individual comparisons are %directly comparable and 
not affected by changes in overhead or variance between workflow runs.

\textbf{(2) Scaling across cluster sizes.}
We execute the same workflows on 2, 3, 4, and 8 nodes using identical Nextflow configurations.
Using the same experimental methodology as in experiment (1), we examine how the attribution accuracy of Nf-PEAK changes for different cluster sizes. %, again using RAPL as ground truth.
Note that it is not sufficient to compare the results produced by Nf-PEAK across cluster sizes, since the differences in available computational resources cause changes in task scheduling and parallelism.
Therefore, the same workflow with the same configuration and input data has a different runtime and energy consumption %of individual tasks and the workflow as a whole 
for different cluster sizes.
%This experiment is not intended to show that workflows have constant energy across node counts (they do not, because parallelism and runtime change), but to test whether Nf-PEAK remains \emph{plausible} and stable when the distributed execution changes.

\textbf{(3) Co-located CPU load.}
To emulate co-located workloads, we run \texttt{stress-ng}\footnote{https://manpages.ubuntu.com/manpages/jammy/man1/stress-ng.1.html, last accessed: January 12, 2026} CPU stressors on \(n\in\{1,2,4,6,8\}\) hardware threads on 4 nodes during workflow execution.
We choose a maximum of 8 loaded threads (of 32) so that at least 24 threads remain available.
Since none of the tested workflows uses more than 24 threads for a single task, %in our setup, this keeps 
workflows remain executable, enabling evaluation of performance under co-located workloads. %configurations comparable while adding substantial unrelated load that influences RAPL values.

\textbf{The mean absolute percentage error (MAPE)} quantifies the deviation between the workflow energy attributed by Nf-PEAK and a reference value based on RAPL.
Let \(E_{\text{RAPL}}\) be the total CPU+DRAM energy measured by RAPL on all nodes during a workflow run. %(summing Package and DRAM domains).
Let \(E_{\text{static}}\) be the static CPU and DRAM energy the same nodes would consume while idle for the same duration, estimated from a pre-run idle-power measurement as described in Section~\ref{Attribution}.
Because Nf-PEAK already attributes part of the static energy to workflow processes, we add only the remaining unattributed static energy before comparing against RAPL:

%\[
%\begin{equation}
%MAPE = 100 \cdot \left|1 - \frac{AE_{\text{PEAK}} + \left(E_{\text{static}}-AE_{\text{static}}\right)}{E_{\text{RAPL}}\cdot E_{\text{Overhead}}}\right|
%\end{equation}
%\]
\begin{equation}
\begin{aligned}
E_{\text{cmp}} &= A_{\text{PEAK}} + \left(E_{\text{static}} - A_{\text{static}}\right)\\
E_{\text{ref}} &= \beta_{\text{overhead}} \cdot E_{\text{RAPL}}\\
\operatorname{MAPE} &= 100 \cdot
\left|1-\frac{E_{\text{cmp}}}{E_{\text{ref}}}\right|
\end{aligned}
\label{eq:mape}
\end{equation}

Here, \(A_{\text{PEAK}}\) is the total CPU and DRAM energy attributed by Nf-PEAK to workflow tasks, including dynamic energy and the static energy fractions assigned by Eq.~(\ref{eq4}) and Eq.~(\ref{eq6}).
\(A_{\text{static}}\) denotes the subset of \(A_{\text{PEAK}}\) that corresponds to the attributed static energy.
\(E_{\text{cmp}}\) is therefore the workflow estimate based on Nf-PEAK after adding the remaining unattributed static energy.
\(E_{\text{ref}}\) is the reference value based on RAPL after discounting the average non-workflow overhead.
We measured \(\beta_{\text{overhead}}=0.97\) for our experimental setup by running an empty workflow that only executes \texttt{sleep}, corresponding to an average overhead of 3\%.
A MAPE value close to zero indicates that the corrected Nf-PEAK estimate closely matches the reference based on RAPL.

\subsection{Experimental Results}\label{ExperimentalResults}
\begin{figure*}[t]
\centering
\begin{minipage}[t]{0.32\textwidth}
  \centering
  \maybeincludegraphics[width=\linewidth, trim={0.0cm 0cm 0.0cm 0.0cm}]{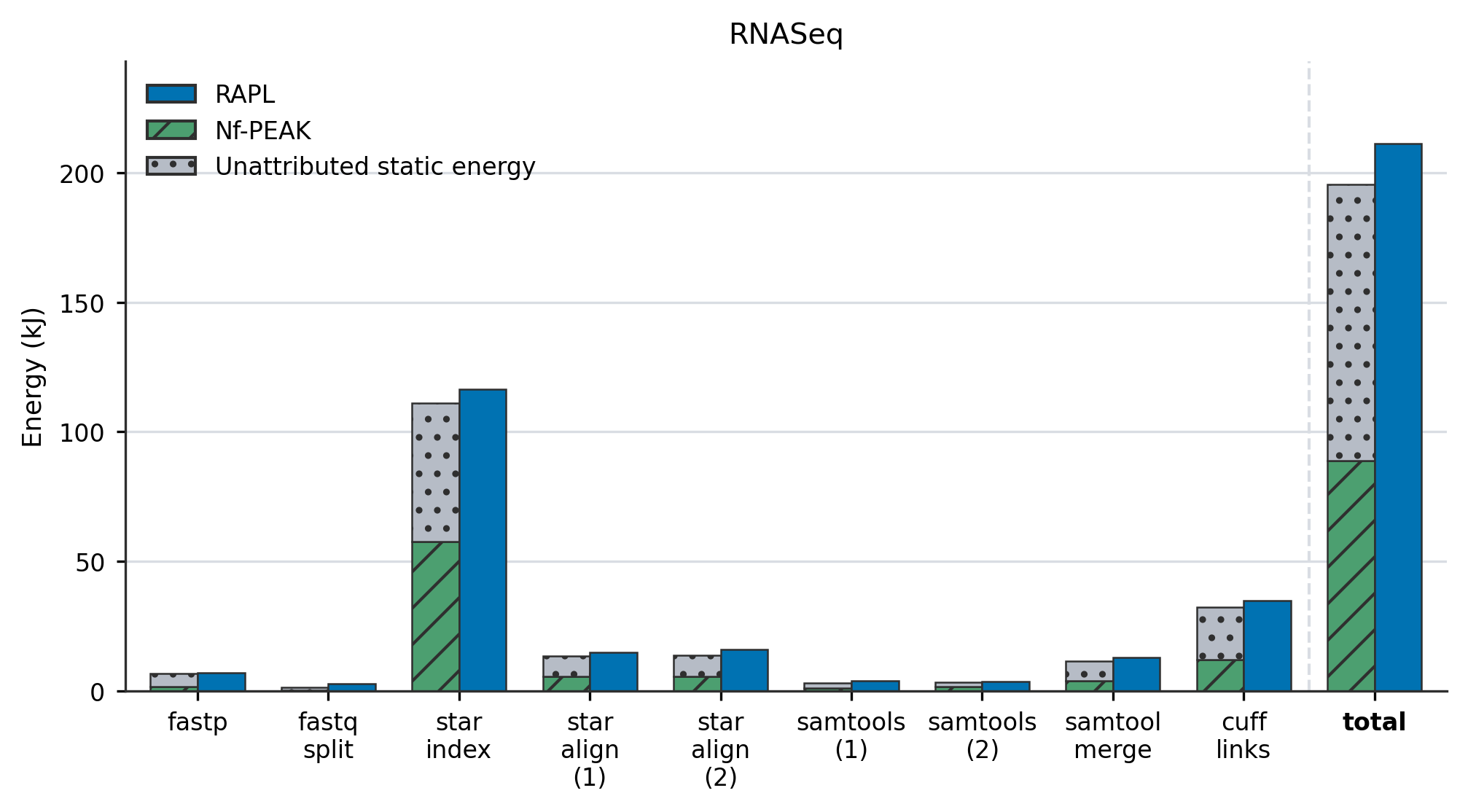}\\[-2pt]
  {\small (a) RNASeq (per-task breakdown with unattributed static energy)}
\end{minipage}%\hfill
%\begin{minipage}[t]{0.49\textwidth}
%  \centering
%  \maybeincludegraphics[width=\linewidth]{task_energy_summary_pkg_energy_hist.png}\\[-2pt]
%  {\small (b) Sarek (histogram of energy consumption of all physical tasks)}
%\end{minipage}\hfill
%\par
%\vspace{5mm}
\begin{minipage}[t]{0.32\textwidth}
  \centering
  \maybeincludegraphics[width=\linewidth]{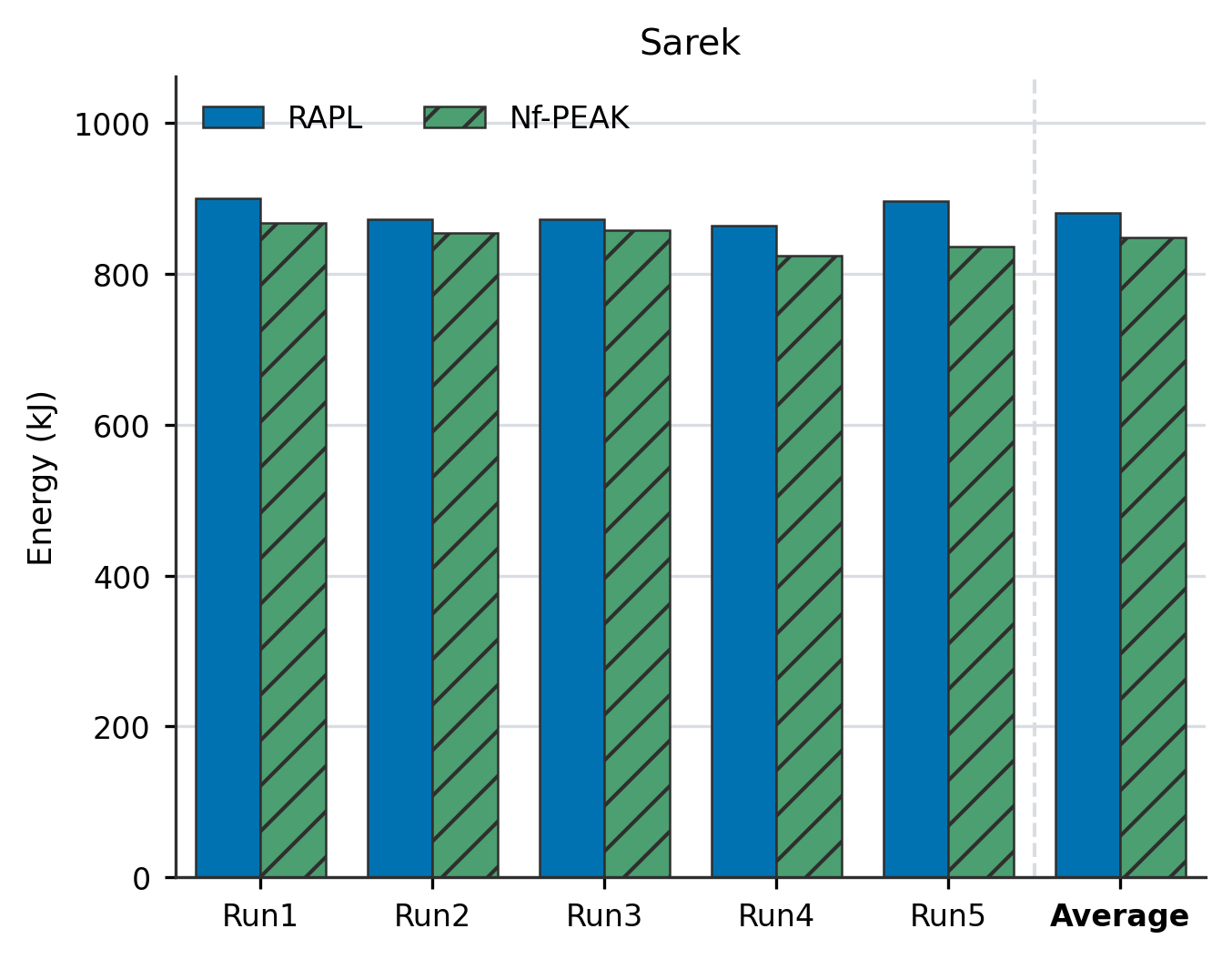}\\[-2pt]
  {\small (b) Sarek (aggregated energy over five %individual 
  runs showing a close match with RAPL% despite variance between runs
  )}
\end{minipage}\hfill
\begin{minipage}[t]{0.32\textwidth}
  \centering
  \maybeincludegraphics[width=\linewidth]{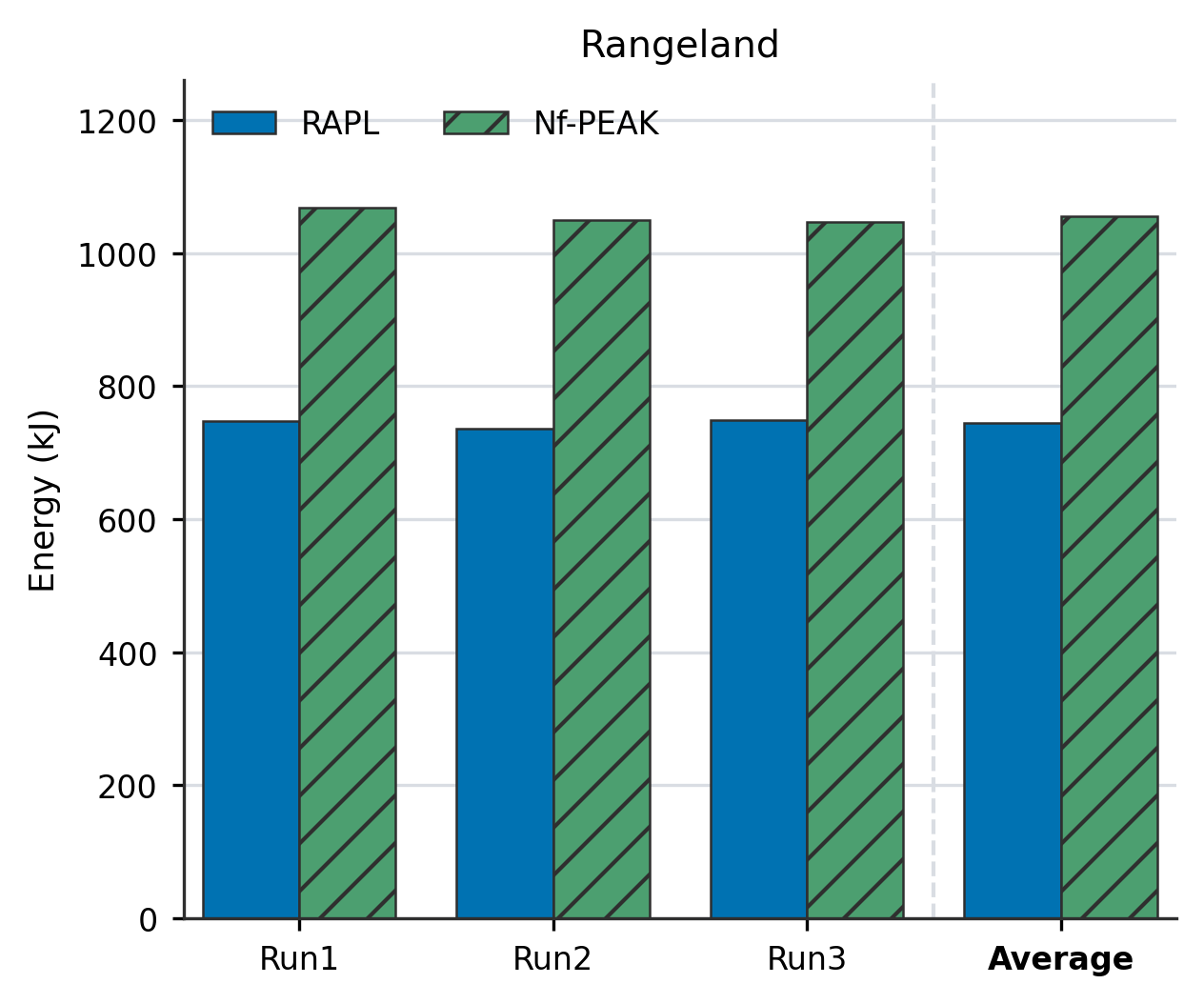}\\[-2pt]
  {\small (c) Rangeland (aggregated energy over three individual runs showing over-attribution)}
\end{minipage}
\caption{Isolated results on 2 nodes for RAPL and Nf-PEAK. %: total energy measured by RAPL and energy attributed by Nf-PEAK. 
For RNASeq, unattributed static energy is shown separately to highlight that %static energy is not fully attributable when resources remain unused and 
the combination of attributed energy and unattributed static energy match the energy measured with RAPL.}
\label{fig:isolated_bars_main}
\end{figure*}

%\subsubsection*{Correlation with RAPL in isolated runs}
\textbf{Correlation with RAPL in isolated runs.} 
Figure~\ref{fig:isolated_bars_main} summarizes isolated 2-node results.
Since the workflows are executed in isolation with no other workloads on the node, we expect the total energy attributed by Nf-PEAK to the workflow to be equal to the dynamic energy of the nodes, as measured by RAPL during execution, plus the fraction of static energy attributed to the tasks, depending on their CPU time. When we add the remaining static energy that we previously determined using RAPL to the energy attributed by Nf-PEAK and account for overhead, the result should match the energy measured with RAPL during execution. %If this was the case, the resulting MAPE of the experiment would be 0 since it would perfectly match the expected result. If there is a discrepancy between the results expected based on RAPL and reported by Nf-PEAK, the MAPE reports the deviation in percent.

For RNASeq, Nf-PEAK attributes only a fraction of total RAPL energy, because most tasks run on few CPU cores and there is no parallel task execution. The remaining energy that has not been attributed by Nf-PEAK consists of static energy unassigned due to low CPU time of the workflow tasks and overhead.
Because RNASeq tasks do not overlap on a node in our configuration, we can compare per-task RAPL deltas directly with the energy attributed to each task.
Across the entire workflow, the energy attributed by Nf-PEAK %added to the remaining, non-attributed static energy and overhead, accounts for 95.4\% of the energy measured with RAPL, resulting 
results in \(4.6\%~MAPE\).

Investigating the individual tasks of the RNASeq workflow, as visible in Figure \ref{fig:isolated_bars_main}a, reveals that %the energy attributed by 
Nf-PEAK is accurate across all tasks in the workflow. For each task, the attributed energy %is slightly lower than what would be expected based on RAPL, resulting 
results in a single-digit MAPE. %for every task. 
This result confirms that Nf-PEAK is capable of accurately attributing energy across a range of different workflow tasks, including tasks with low CPU usage, such as \texttt{samtools}, and %such that on average use more than 13 of the available 16 physical CPU cores available on our nodes
tasks with high CPU usage, such as \texttt{star\_index}.

For Sarek, the high sustained load on the nodes caused by task parallelism leads to attribution values that are close to the total energy used during workflow execution as measured with RAPL (Figure \ref{fig:isolated_bars_main}b).
%After accounting for static energy and overhead, Nf-PEAK attributes 99.2\% of the energy expected based on RAPL in isolated runs (\(MAPE=0.8\%\)). 
Accounting for static energy and overhead, Nf-PEAK achieves \(0.8\%~MAPE\), showing that Nf-PEAK is capable of attributing energy accurately in scenarios with parallel tasks and high CPU utilization. %, confirming that Nf-PEAK can be used to attribute energy for compute-intensive, highly parallel workflows. 
%Notably, there was no over-attribution of energy despite the cluster frequently running near its maximum CPU capacity (including multi-threading) during the execution of Sarek. Over multiple individual runs, energy attributed by Nf-PEAK always was slightly below the value measured with RAPL, despite changes in total energy consumption between runs due to scheduling differences (Figure \ref{fig:isolated_bars_main}b).

%For highly parallel workflows such as Sarek, an evaluation of the exact energy attributed to individual physical tasks has limited value due to the higher variance between runs that occurs naturally due to differences in scheduling. Nevertheless, the task-based attribution of Nf-PEAK allows to draw conclusions about the structure of the workflow and the resources used by its individual logical and physical tasks. Figure \ref{fig:isolated_bars_main}b is a histogram of the energy consumed by all physical tasks of Sarek as attributed by Nf-PEAK. It shows that Sarek contains many less energy-intensive tasks with an energy consumption between 3,000J and 6,000J, but only few task with very low energy consumption. Additionally, there are individual outliers that have a higher energy consumption of up to 40,000J. A more detailed analysis of the distribution of energy between individual logical and physical tasks in Sarek can be found in Appendix \ref{app:Task}.

Rangeland contains many sub-second and few-second physical tasks and therefore represents the main failure mode of Nf-PEAK. %Polling-based discovery and interval-based sampling like those used in Nf-PEAK can miss parts of their execution, leading to inaccurate energy attribution %when the polling frequency is too low to collect 
%due to insufficient process- and RAPL-data.
Polling-based discovery and interval-based sampling observe task start, process lifetime, and RAPL energy deltas in different sampling windows.
For very short-lived processes, the resulting process statistics are insufficient to accurately determine which process should receive which part of an interval's energy.
In our experiments, this limitation of %the attribution strategy implemented in 
Nf-PEAK leads to a high attribution error of \(46.1\%~MAPE\),
%on two nodes, remaining stable across repeated runs 
as shown in Figure \ref{fig:isolated_bars_main}c.
Note that the energy attributed to Rangeland by Nf-PEAK is higher than the total energy measured by RAPL, which is physically impossible. This is because energy is attributed for each process individually during runtime and then aggregated to tasks. If over-attribution occurs for multiple processes, the result can be an over-attribution for the total energy used. Such an error can be detected after workflow execution, %since the sum of all attributions for parallel tasks exceeds the total energy measured with RAPL. However, 
but post-run correction is difficult, because it is not known which attributions are inaccurate. %tasks of the workflow were attributed too much energy.
%This result with a much higher error than the other experiments 
The high attribution error of Nf-PEAK on Rangeland highlights the limitations of the used polling- and interval-based attribution strategy. %used by Nf-PEAK. 
If the workflow contains multiple short tasks, as Rangeland does with over 2000 tasks with a length of less than ten seconds in our experiments, attributions for individual physical tasks are less precise, reducing overall accuracy significantly. %and small errors for individual tasks accumulate to produce 
%producing high overall errors. %in the result. 
%The impact of this effect is dependent on the hyperparameters of Nf-PEAK (polling-interval and interval for RAPL reads). However, using smaller intervals also creates higher overhead by Nf-PEAK. A more detailed discussion about the tradeoff between overhead and accuracy for short tasks can be found in Section \ref{ExperimentalResults} below.

%\begin{figure*}[t]
%\centering
%\begin{minipage}[t]{0.32\textwidth}
%  \centering
%  \maybeincludegraphics[width=\linewidth]{RNASeq_2nodes_bar_switched_updated_new.png}\\[-2pt]
%  {\small (a) RNASeq (per-task breakdown)}
%\end{minipage}\hfill
%\begin{minipage}[t]{0.32\textwidth}
%  \centering
%  \maybeincludegraphics[width=\linewidth]{Sarek_Nf-PEAK_bar_new.png}\\[-2pt]
%  {\small (b) Sarek (aggregated over individual runs)}
%\end{minipage}\hfill
%\begin{minipage}[t]{0.32\textwidth}
%  \centering
%  \maybeincludegraphics[width=\linewidth]{Rangeland_Nf-PEAK_bar_new.png}\\[-2pt]
%  {\small (c) Rangeland (aggregated over individual runs)}
%\end{minipage}
%\caption{Isolated workflow-level results on two nodes: total energy measured by RAPL and energy attributed by Nf-PEAK. For RNASeq, unassigned static energy is shown to highlight that static energy is not fully attributable when resources remain unused.}
%\label{fig:isolated_bars_main}
%\end{figure*}

%\subsubsection*{Scaling across cluster sizes}
\textbf{Scaling across cluster sizes.} 
To examine how Nf-PEAK behaves when the cluster size increases, we repeated the same set of experiments that we previously conducted on 2 nodes, using 3, 4, and 8 nodes of the same cluster. %The workflows were executed in the same configurations and on the same input data. 
%Therefore, we 
We expect the overall energy consumption of the workflows to be similar to that on 2 nodes, with any differences mainly being caused by runtime differences due to the execution of more tasks in parallel rather than sequentially. %Any changes in energy consumption due to larger cluster size will also reflect in RAPL, which we continue to use as a basis to compute the expected attributable energy.
%Figure~\ref{fig:scaling_nodes} and Table~\ref{tab:scaling_nodes} 
%Figure~\ref{fig:scaling_nodes} summarizes energy consumed when executing the workflows on three and four nodes in addition to the results from our first set of experiments on two nodes.

RNASeq does not make use of more resources than are available already on 2 nodes in the configuration used in our experiments, resulting in very similar runtime and energy consumption. The energy attributed by Nf-PEAK is very stable as well, resulting in an error of \(0.8\%~MAPE\) for an execution on 4 nodes and \(4.1\%~MAPE\) for an execution on 8 nodes on average. %. Together with slightly higher overhead due to more involved nodes, which slightly decreases the amount of energy expected to be attributed by Nf-PEAK, this leads to an error of \(MAPE=0.8\%\) for an execution on four nodes.

Unlike RNASeq, Sarek includes parallel workflow tasks and benefits from additional resources. As a result, the runtime on 4 nodes decreases to 84.6\% in comparison to execution on 2 nodes, and attributed energy decreases to 82.3\%. The resulting error remains low at \(4.2\%~MAPE\). On 8 nodes, further reductions in parallel task execution lead to a low error of \(0.8\%~MAPE\). %Compared to the experiment on two nodes, the deviation from the theoretically attributable energy of an execution on four nodes is slightly higher, but it remains overall low at \(MAPE=4.2\%\).

\begin{figure*}[t]
\centering
\begin{minipage}[t]{0.32\textwidth}
  \centering
  \maybeincludegraphics[width=\linewidth]{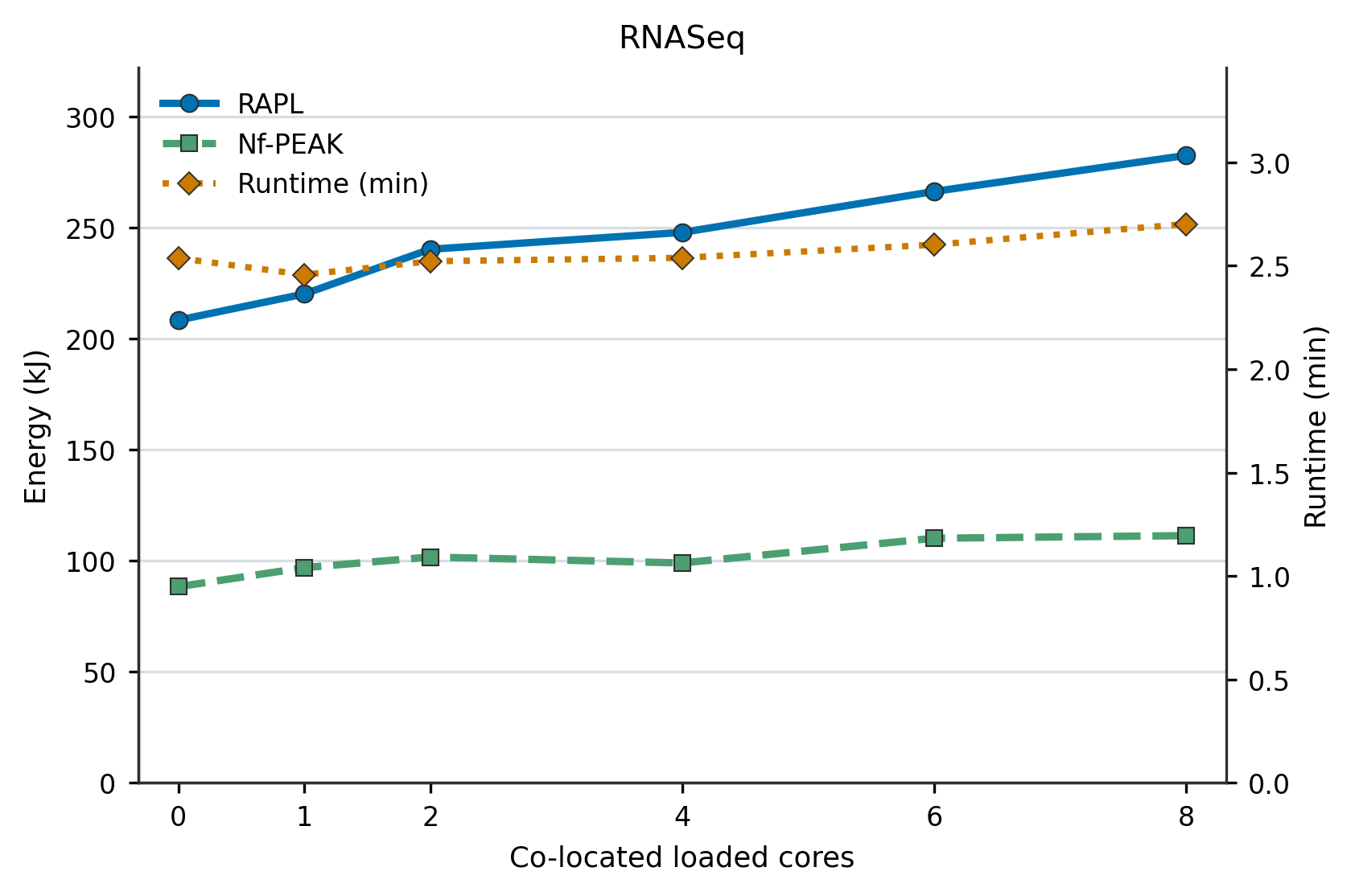}\\[-2pt]
  {\small (a) RNASeq}
\end{minipage}\hfill
\begin{minipage}[t]{0.32\textwidth}
  \centering
  \maybeincludegraphics[width=\linewidth]{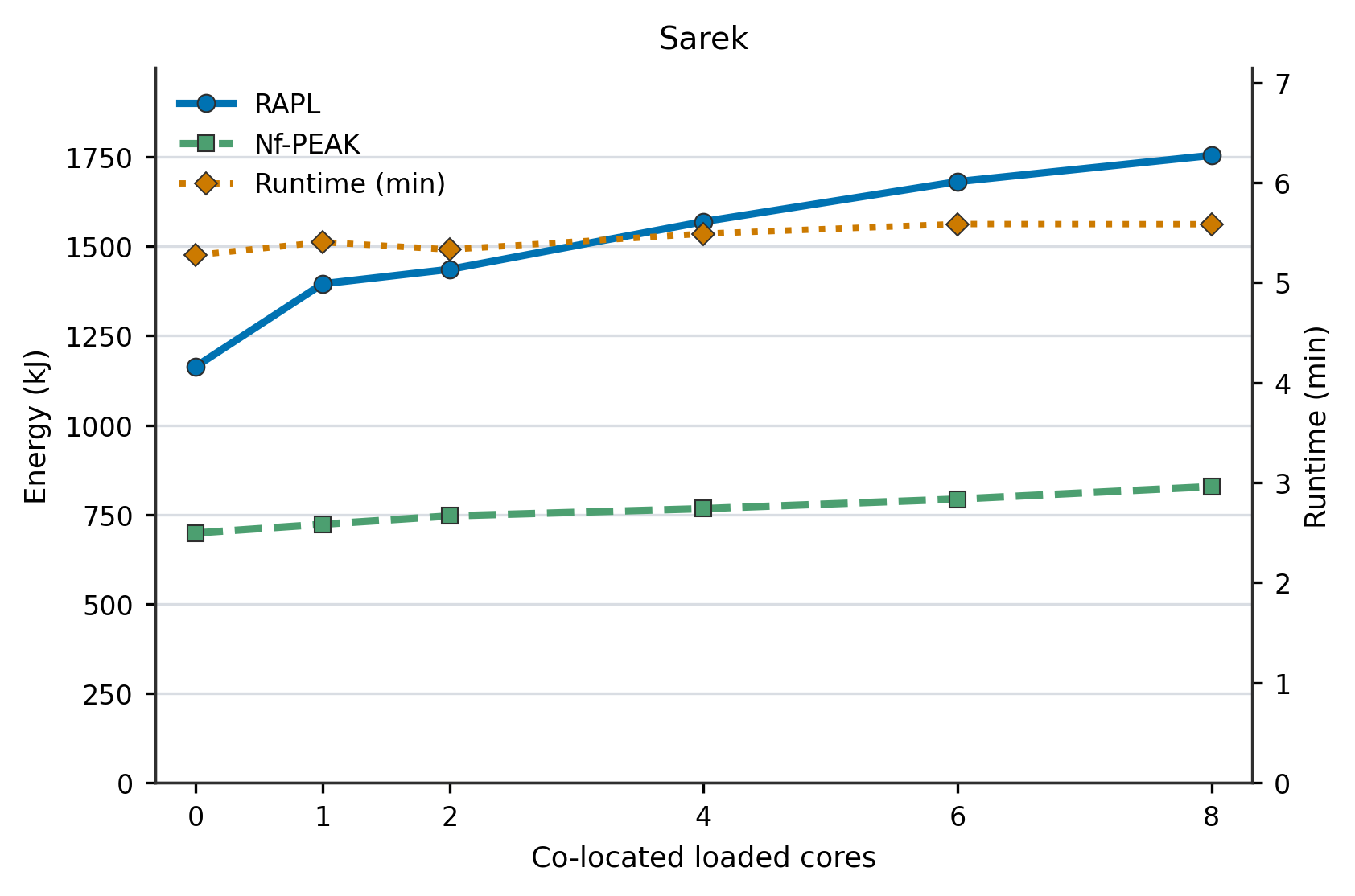}\\[-2pt]
  {\small (b) Sarek}
\end{minipage}\hfill
\begin{minipage}[t]{0.32\textwidth}
  \centering
  \maybeincludegraphics[width=\linewidth]{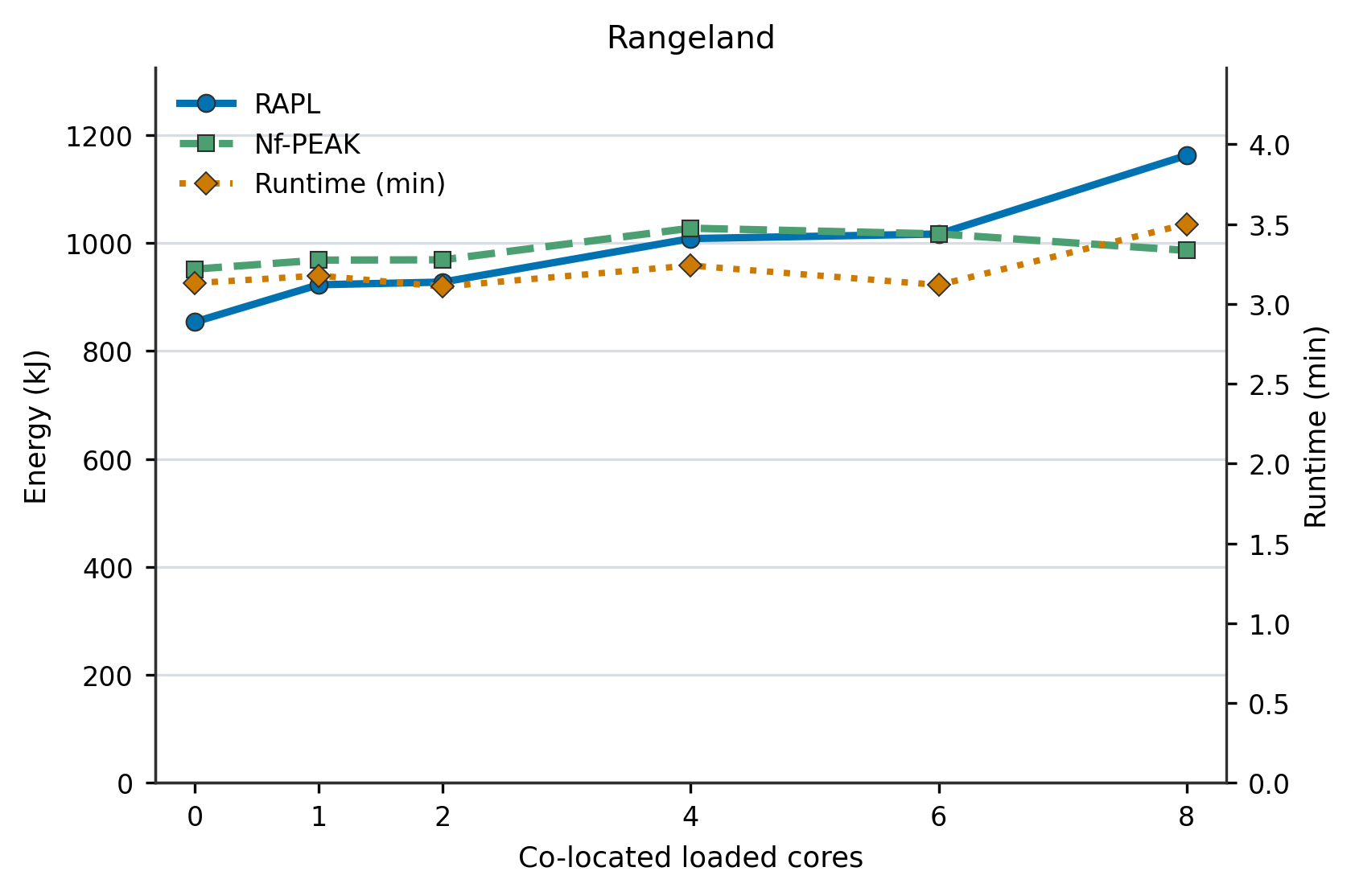}\\[-2pt]
  {\small (c) Rangeland}
\end{minipage}
\caption{Workflow-level energy %(RAPL vs.\ Nf-PEAK) 
and runtime on 4 nodes under increasing co-located CPU load. %(number of stressed hardware threads). 
Energy measured by RAPL increases with the %number of cores under 
co-located load for all workflows. Meanwhile, the tested workflows are only slightly affected by the co-located load, as indicated by the mostly stable runtimes. Energy attributed by Nf-PEAK %does not follow the increase by RAPL and 
remains stable as well, %with only slight increases under high co-located load, 
indicating that Nf-PEAK is capable of dealing with co-located load during energy attribution.}
\label{fig:load_plots_main}
\end{figure*}

Rangeland scales well across multiple nodes. When executed on 4 nodes, its runtime is equal to 64.8\% of the runtime on 2 nodes. At the same time, energy attributed by Nf-PEAK also decreases, resulting in a lower error of \(14.9\%~MAPE\) for 4 nodes and \(12.4\%~MAPE\) for 8 nodes. %since the total energy used by Rangeland according to RAPL increases, decreasing the over-attribution by Nf-PEAK. However, the significantly larger MAPE compared to the other tested workflows still highlights the general problem of Nf-PEAK with attributing energy for very short workflow tasks.
While the error of Nf-PEAK for Rangeland on 4 and 8 nodes is smaller due to reduced over-attribution, the MAPE remains significantly larger compared to RNASeq and Sarek, highlighting the problem of attributing energy for short tasks.
\textbf{Behavior under co-located CPU load.} 
Our third experiment aims to quantify the accuracy of Nf-PEAK under co-located, unrelated CPU load. %not related to the workflow.
While executing each workflow on 4 and 8 nodes using the same configuration and input data as in the second experiment, we execute \(n\in\{1,2,4,6,8\}\) hardware threads of a \texttt{stress-ng} CPU benchmark, fully loading one logical CPU core each. We expect the impact on individual workflow tasks to be small, because none of the tested tasks %workflows contains a single task that uses 
use more than the 24 remaining unoccupied logical cores of our CPUs. Consequently, the energy attributed to each workflow by Nf-PEAK should be nearly the same. %, with differences only caused by slightly longer runtime because less tasks can be scheduled in parallel.
Figure~\ref{fig:load_plots_main} shows total energy as measured by RAPL, energy attributed by Nf-PEAK, %to each workflow 
and workflow runtime under increasing co-located CPU load %induced by \texttt{stress-ng} 
on 4 nodes.
As expected, total RAPL energy increases with additional load because previously idle threads now consume dynamic energy.
Meanwhile, the runtime and attributed energy for the workflows themselves remain stable, with moderate increases at higher load levels due to increased runtime, matching our expectations.

For RNASeq, total RAPL energy increases by 35.6\% when 8 threads are loaded compared to no additional load, while runtime increases by 6.5\% and Nf-PEAK attributed energy increases by 26\%.
The increase in attributed energy results in %a slight over-attribution, with an error of 
\(6.4\%~MAPE\). While this value indicates a larger error than in the experiments without co-located load, the attribution of Nf-PEAK remains relatively stable with a limited error even when a quarter of all hardware threads are fully used by an unrelated, but co-located workload. On 8 nodes, the error in attributed energy is again lower with \(2.1\%~MAPE\).

For Sarek, RAPL energy increases by 50.7\%, runtime by 5.8\%, and attributed energy by 18.6\% in an execution on 4 nodes with co-located load.
Similar to RNASeq, the resulting \(7.3\%~MAPE\) %Mean Absolute Percentage Error 
is slightly higher than in scenarios without co-located workload. %In the experiment with co-located load on eight cores, Nf-PEAK achieves \(MAPE=7.3\%\).
This is also true on 8 nodes, where the error is \(4.8\%~MAPE\).

The attributed energy of Rangeland changes by only 3.6\%, while runtime increases by 11.8\% and total RAPL energy by 36.1\%.
This results in a similar increase in MAPE as for the other workflows. With \(19.1\%~MAPE\) on 4 nodes and \(26.2\%~MAPE\) on 8 nodes, the error remains significantly higher than for the other workflows we tested. %, highlighting the issue of short tasks once again.

%Overall, MAPE increases under heavy co-located load for all workflows, but remains below 8\% for RNASeq and Sarek and below 20\% for Rangeland on four nodes.

%\subsubsection*{Impact of monitoring intervals}
\textbf{Impact of monitoring intervals.} 
%Nf-PEAK trades off \emph{coverage} of short-lived tasks against \emph{monitoring overhead}.
Shorter polling and sampling intervals make it more likely to collect sufficient data for accurate energy attribution, %for very short tasks, 
but they also increase the overhead of each individual per-process monitor.

We evaluate this trade-off by gradually shortening both the interval of the polling loop and the interval for reading RAPL energy counters over multiple, otherwise identical workflow runs.
%Figure~\ref{fig:intervals_main} shows experimental results on four nodes for Rangeland, which contains many sub-second tasks and is therefore the only one of the three tested algorithms where shorter monitoring intervals have a measurable impact on the total energy attributed by Nf-PEAK.
On average, the energy attributed by Nf-PEAK for Rangeland increases by 7.7\% when shortening the polling interval from 10\,s to 0.2\,s and the interval for reading RAPL from 4\,s to 0.05\,s.
%Note that the energy attributed by Nf-PEAK to Rangeland was already too high with the longer intervals used in our experiments. Using a shorter interval between measurements only increases the over-attribution, highlighting that it is not caused by insufficient polling intervals.
Note that there is already an over-attribution by Nf-PEAK to Rangeland with the longer intervals used in our experiments. This result highlights that insufficient polling intervals are not the only cause for the reduced accuracy on short workflow tasks.
For the other workflows, no significant changes in the average attributed energy occur for shorter polling intervals.
This indicates that the longer polling intervals used previously are sufficient to collect the data needed for accurate attribution. %, and finer-grained data does not change the results.
%We conclude from this result that the accuracy of Nf-PEAK measured in our other experiments is not influenced by insufficient polling intervals and any error in attribution on RNASeq and Sarek must be due to the used attribution formula, not missing data.
At the same time, shorter polling intervals increase the overhead. %for measurement on all three workflows as indicated by an increase in measured RAPL energy while workflow execution times are stable.
%For Rangeland, the total energy measured with RAPL increases by 15\% between the longest and the shortest tested polling intervals for an otherwise identical workflow execution as shown in Figure \ref{fig:intervals_main}. The other two workflows show similar increases.
Between the longest and the shortest intervals we tested, the total energy measured with RAPL during workflow execution increases by 15\%. %for Rangeland.
%These results show that shorter measurement intervals are causing a measurable increase in overhead, while not affecting the energy attributed by Nf-PEAK except for workflows with very short tasks. In this special case, there is an impact of the measurement intervals, but shortening them does not increase the accuracy of Nf-PEAK, instead worsening the over-attribution caused by other factors.
We therefore conclude that shorter measurement intervals have no measurable advantage for the accuracy of Nf-PEAK on any of the tested workflows, but significantly increase the overhead. %measurement overhead independent of the workflow examined.
%Shorter intervals increase the energy measured with RAPL (overhead), while also slightly increasing the energy attributed to the workflow (more short tasks captured).
%For workflows dominated by longer-running tasks (RNASeq and Sarek), we observed no noticeable increase in attributed energy when shortening intervals, but still an increase in overhead.
%Based on these results, we conclude that the 
%Task-discovery interval of 5\,s and RAPL sampling interval of 2\,s %in our experiments.
%These intervals 
Task-discovery and RAPL sampling intervals of 5\,s and 2\,s, respectively,
are sufficient for accurate attribution %on the tested workflows 
while ensuring low overhead. %keeping the measurement overhead low.

%\begin{figure}[t]
%\centering
%\maybeincludegraphics[width=1.00\linewidth]{Rangeland_configurations_Nf-PEAK_new.png}
%\caption{Rangeland on four nodes under five different monitoring interval configurations: RAPL energy (total) and Nf-PEAK attributed energy.}
%\label{fig:intervals_main}
%\end{figure}

%\subsubsection*{Comparison to Kepler}
\textbf{Comparison to Kepler.} 
Kepler
% ~\cite{amaral_kepler_2023}
is a widely used Kubernetes energy exporter that also provides energy attribution on container/task-level granularity.
For task-level energy attribution, Kepler uses a linear CPU-time heuristic\footnote{https://sustainable-computing.io/archive/design/power\_model/, last accessed: February 5, 2026}:
%\[
\begin{equation}
\textit{Workload Power} = \left(\frac{\Delta \textit{Workload CPU Time}}{\Delta \textit{Node CPU Time}}\right)\cdot \textit{Power}
\end{equation}
%\]
%Table~\ref{tab:kepler_abs_main} lists absolute energy values (Joule) for RAPL, Nf-PEAK, and Kepler in isolated runs on four nodes.
%RAPL reports total node energy (including static energy and platform overhead), whereas Nf-PEAK and Kepler report energy attributed to workflow tasks. For our experiments, we used Kepler v0.8.0.% in revision bfae0a73afbd53f895b3be7643eb0c95ecd6e94f.

%\begin{table}[t]
%    \centering
%    \caption{Absolute energy values for RAPL, Nf-PEAK and Kepler (isolated runs on four nodes, averaged over three runs).}
%    \label{tab:kepler_abs_main}
%    \begin{tabular}{lccc}
%        \toprule
%        \textbf{Workflow} & \textbf{RAPL} & \textbf{Nf-PEAK} & \textbf{Kepler} \\
%        \midrule
%        RNASeq      & 208,422\,J  & 88,336\,J  & 114,021\,J \\
%        Sarek       & 1,163,911\,J & 698,462\,J & 701,274\,J \\
%        Rangeland   & 853,844\,J  & 951,734\,J & 540,400\,J \\
%        \bottomrule
%    \end{tabular}
%\end{table}

To enable a direct comparison of the accuracy of both approaches, we ran Kepler's energy attribution alongside that of Nf-PEAK for all three tested workflows, with and without co-located load, on 4 nodes. 
For our experiments, we used Kepler v0.8.0.
%To compare accuracy across workflows and scenarios, 
Table~\ref{tab:PE_Comparison} summarizes MAPE for both approaches in isolated runs and under load on 4 nodes.
Across the three workflows, Nf-PEAK achieves an average MAPE of 6.6\% without additional load and 10.9\% under load.
In comparison, Kepler achieves an average MAPE of 17.4\% without load and 22.5\% with load. %eight logical cores under load, both significantly worse than Nf-PEAK.
Our results indicate that Nf-PEAK on average outperforms Kepler without and with co-located load.
% Even under heavy co-located load %better on average than 
% Kepler in scenarios without load on average.
Kepler performs competitively in some tests, but degrades more under co-located load for Sarek and Rangeland in our experiments.

Kepler's energy attribution model for physical tasks is based on a linear function with CPU-time share as input.
When unrelated co-located load increases the node-level CPU time and RAPL energy, this changes the denominator used for attribution and reduces the energy attributed to a physical task, even if the physical task itself is unchanged.
Nf-PEAK also uses process-level CPU time and memory usage, but transforms them through a non-linear credit model before assigning dynamic energy, which reduces the sensitivity to changes in aggregate node utilization.
%Figure~\ref{fig:Kepler_Sarek_Load} illustrates this behavior for Sarek: while Nf-PEAK slightly increases attributed energy with higher load, slightly degrading the result, Kepler's linear CPU-time model attributes \emph{less} workflow energy as load increases. This behavior moves the attribution provided by Kepler further away from the expected value with increasing load and causes a significant increase in MAPE.

%\begin{table*}[t]
%    \centering
%    \caption{Mean Absolute Percentage Error (MAPE) for Nf-PEAK and Kepler in isolated runs (!L) and under load (L: eight stressed threads), averaged over three runs on four nodes. Lower is better. The Deviation column shows the deviation of the respective attribution tool under load compared to an isolated run in percentage points.}
%    \label{tab:PE_Comparison}
%    \begin{tabular}{lcc|cccc}
%        \toprule
%        \textbf{Workflow} & \textbf{Nf-PEAK(!L)} & \textbf{Kepler(!L)} & \textbf{Nf-PEAK(L)} & \textbf{Deviation} & \textbf{Kepler(L)} & \textbf{Deviation} \\
%        \midrule
%        RNASeq      & \textbf{0.8\%} & 13.5\% & 6.4\% & \textbf{5.6~pp} & \textbf{3.5\%} & -10.0~pp \\
%        Sarek       & 4.2\% & \textbf{4.0\%} & \textbf{7.3\%} & \textbf{3.1~pp} & 16.7\% & 12.7~pp \\
%        Rangeland   & \textbf{14.9\%} & 34.8\% & \textbf{19.1\%} & \textbf{4.2~pp} & 47.4\% & 12.6~pp \\
%        \bottomrule
%    \end{tabular}
%\end{table*}

\begin{table}[t]
    \centering
    %\footnotesize
    %\setlength{\tabcolsep}{3pt}
    %\renewcommand{\arraystretch}{1.05}
    \caption{%Mean Absolute Percentage Error (MAPE) 
    MAPE for Nf-PEAK and Kepler in isolated runs %(No Load) 
    and under load %(Load: eight stressed threads) 
    (8 threads) on 4 nodes. Lower is better. Deviation denotes the signed deviation under load relative to the isolated result in percentage points.}
    \label{tab:PE_Comparison}
    \begin{tabular}{llccc}
        \toprule
        \textbf{Workflow} & \textbf{Tool} & \textbf{No Load} & \textbf{Load} & \textbf{Deviation} \\
        \midrule
        \multirow{2}{*}{RNASeq}
          & Nf-PEAK & \textbf{0.8\%}  & 6.4\%           & \textbf{5.6 pp} \\
          & Kepler  & 13.5\%          & \textbf{3.5\%}  & -10.0 pp        \\
        \addlinespace[0.2em]
        \multirow{2}{*}{Sarek}
          & Nf-PEAK & 4.2\%           & \textbf{7.3\%}  & \textbf{3.1 pp} \\
          & Kepler  & \textbf{4.0\%}  & 16.7\%          & 12.7 pp         \\
        \addlinespace[0.2em]
        \multirow{2}{*}{Rangeland}
          & Nf-PEAK & \textbf{14.9\%} & \textbf{19.1\%} & \textbf{4.2 pp} \\
          & Kepler  & 34.8\%          & 47.4\%          & 12.6 pp         \\
        \bottomrule
    \end{tabular}
\end{table}

%\begin{figure}[t]
%\centering
%\maybeincludegraphics[width=1.00\linewidth]{Kepler_Sarek_Load_Nf-PEAK_new.png}
%\caption{Energy for Sarek on four nodes under increasing co-located CPU load: RAPL (total), Nf-PEAK, and Kepler. Runtime is shown for context.}
%\label{fig:Kepler_Sarek_Load}
%\end{figure}

%\subsubsection*{Task-Level Analysis for Parallel Workflows}\label{sec:Task}
\textbf{Task-Level Analysis for Parallel Workflows.} 
The task-level results produced by Nf-PEAK allow insights into the distribution of energy consumption even for highly parallel scientific workflows. %, where the accuracy of the attribution can not be directly confirmed using RAPL.

%Figure \ref{fig:bar-logical} shows the energy consumption of each logical task in the Sarek workflow as a bar chart.
%The figure shows that not all tasks contribute equally to energy consumption.
%In the case of Sarek, only five of the fifteen logical tasks contribute a large amount of energy (more than 7500J), while the other ten logical tasks contribute less than 2500J each (except for T15, which consumes slightly over 2500J).
%From this diagram it can be concluded that a future optimization of the workflows should focus on the five tasks with high energy consumption (T03, T04, T05, T09 and T11).
%Optimizing the other tasks can only gain minor improvements in total energy consumption, because their contribution is small in the first place.

\begin{figure}[t]
\centering
\maybeincludegraphics[width=0.89\linewidth, trim={0.0cm 0.0cm 0.0cm 0.0cm},clip]{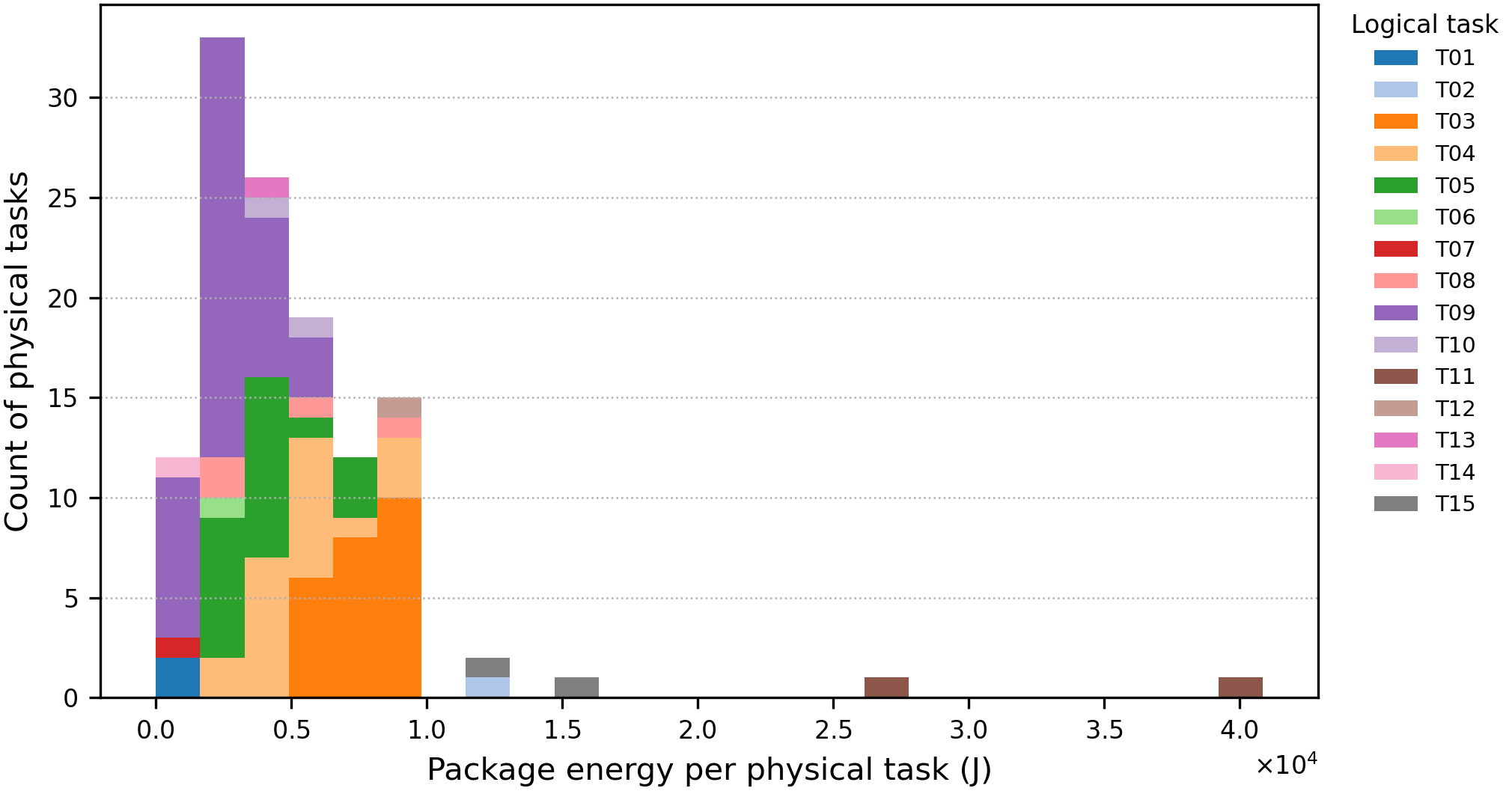}
\caption{Histogram of the energy consumption of each physical task in the Sarek workflow. Task types of each physical task are marked by colors.}% For the associated names used by the workflow engine for each task, refer to Table \ref{tab:Task-names}.}
\label{fig:hist-physical}
\end{figure}

%Figure \ref{fig:hist-physical} shows the distribution of energy consumption for Sarek's physical tasks, as well as the corresponding logical tasks.
Figure \ref{fig:hist-physical} shows the distribution of energy consumption for Sarek's physical tasks and groups them by logical task.
%The colors show the logical task each physical task belongs to.
It highlights that the total energy consumption of logical tasks is not always correlated with the energy consumption of individual physical tasks.
While T9 consumes about 12,000\,J of energy and contains many physical tasks, most of these tasks use relatively little energy.
Meanwhile, T11 consumes less total energy (7,500\,J), but contains only 2 physical tasks with much higher energy consumption.

To identify opportunities for optimization in individual workflow tasks and task scheduling, it is important to consult data regarding both the logical workflow tasks as well as individual physical tasks.
For both categories of tasks, energy consumption can be analyzed using Nf-PEAK data, making it useful not only for reporting energy consumption, but also for identifying opportunities for workflow optimization.

\section{Discussion}\label{Discussion}
%\subsection*{Accuracy and robustness}
\textbf{Accuracy and robustness.} 
Across three real workflows from two scientific domains, Nf-PEAK produces accurate task-level energy estimates, outperforming Kepler on average.
It remains stable across cluster sizes and under co-located load, showing a smaller increase in deviation from the expected result than Kepler.% when executed under heavy co-located workload.
%The main benefit stems from modeling non-linear CPU power behavior via the \(\gamma\) exponent, which reduces systematic bias compared to linear CPU-time heuristics in underutilized regimes.

%\subsection*{Deployment}
\textbf{Deployment and security.} 
Nf-PEAK is designed for managed clusters where users cannot install node software.
%The containerized deployment keeps the workflow execution environment unchanged, but monitoring requires access to host information regarding processes and energy.
%In practice, access may be restricted by cluster administrators in multi-tenant environments.
The workflow pods are executed unchanged, but the monitoring pods require host-level visibility.
Concretely, they use \texttt{hostPID: true} to inspect host processes and a read-only mount of the host \texttt{powercap} interface to read RAPL counters.
These privileges weaken the isolation normally provided by Kubernetes.
Thus, Nf-PEAK should be treated as an administrator-approved observability component, not as an ordinary user workload.
Clusters that enforce the Baseline or Restricted Kubernetes Pod Security Standards disallow host namespaces and hostPath volumes\footnote{https://kubernetes.io/docs/concepts/security/pod-security-standards/, last accessed: May 12, 2026}; such clusters require a policy exception or an alternative administrator-provided telemetry interface.
%In comparison, Kepler is also deployed in containers and requires access to the same information, which it reads through eBPF~\cite{amaral_kepler_2023}.
In comparison, Kepler has similar deployment constraints because it also requires host-level visibility, which it obtains through eBPF~\cite{amaral_kepler_2023}.
This method is efficient, but requires access to kernel hooks of the host machine.
%It can also be deployed in containers on Kubernetes.
Kepler produces measurement results for whole nodes as well as attributions for individual tasks/containers.
The installation requirements for Nf-PEAK and Kepler are very similar, except that Kepler requires Go 1.21+\footnote{https://sustainable-computing.io/kepler/installation/guide/\#prerequisites, last accessed: February 27, 2026} to implement its eBPF kernel extensions.

%\subsection*{Limitations}
\textbf{Limitations.} 
The main accuracy limitation of Nf-PEAK is per-physical-task attribution for very short tasks.
Nf-PEAK relies on periodic polling for task discovery and interval-based sampling.
For very short tasks, the collected performance and energy data can be insufficient, resulting in inaccurate attributed energy.
This can be observed in our experiments with the Rangeland workflow, where the energy attributed by Nf-PEAK is almost 50\% higher than expected and even exceeds the total energy measured with RAPL for the involved nodes.
%Shortening the polling intervals for RAPL and performance counters significantly increases the total energy measured with RAPL during workflow execution, by up to 15\%, while not improving the quality of the attributions made by Nf-PEAK on short tasks.
Shortening the task-discovery and RAPL sampling intervals does not remove this limitation in our experiments: it increases the total energy measured during workflow execution by up to 15\%, while not improving attribution quality for short tasks.
We therefore consider Nf-PEAK reliable for workflows whose physical tasks are observed over multiple sampling intervals, but not for accurate per-physical-task attribution in workflows dominated by sub-second or few-second tasks. For the sampling intervals used in this work, energy can not be reliably attributed to tasks with a length of less than 15 seconds.

Furthermore, RAPL measures only CPU and DRAM energy. Additional energy used by storage, network, and accelerators is not covered by our current implementation.
GPU support could be especially relevant, since GPU accelerators are often critical components of modern data-intensive workflows.
The NVIDIA Management Library (NVML), which also underlies \texttt{nvidia-smi}, exposes information about current board power draw, power limits and GPU utilization\footnote{https://developer.nvidia.com/management-library-nvml, last accessed: May 13, 2026}.
An Nf-PEAK extension could add a GPU monitor that maps CUDA processes to Kubernetes pod UIDs, integrates sampled GPU power, and aggregates the resulting GPU energy to the same physical Nextflow tasks as the current attribution based on CPU and DRAM.
Due to a lack of access to compute infrastructure with modern NVIDIA accelerators, such an extension remains future work.

%\subsection*{Threats to validity}
%Our results are based on a single cluster model and a specific software stack (Kubernetes, Docker, Nextflow).

%The cluster used for experiments consists of single-socket nodes. While Nf-PEAK was implemented to take multi-socket designs into account, we were not able to evaluate its accuracy when deployed on such cluster nodes on the hardware available for our experiments.

%The calibrated \(\gamma\) value and the absolute accuracy of RAPL counters can differ across CPU generations and firmware configurations, which may affect attribution quality.

%Moreover, workflow characteristics depend on the chosen input data and pipeline configuration.
%We mitigate this threat by evaluating three diverse, real-world nf-core workflows, repeating runs to expose variance, and reporting both isolated and stressed scenarios. Nevertheless, further validation on additional hardware and workloads is necessary to fully generalize the reported error ranges.

\section{Related Work}\label{RelatedWork}
Energy measurement and attribution have been studied from several angles, including low-level hardware counter access, process/container attribution, and higher-level carbon accounting.

\textbf{Kubernetes energy monitoring and attribution.}
Kepler~\cite{amaral_kepler_2023} is a widely used Kubernetes energy exporter that provides container/workload-level attribution and can additionally train machine-specific models~\cite{amaral_process-based_2024}.
Other Kubernetes-focused tools (e.g., Scaphandre\footnote{https://github.com/hubblo-org/scaphandre, last accessed: March 22, 2026}) expose node energy and attribute it to containers primarily through linear resource shares.
These systems are popular because they integrate with cluster observability stacks (e.g., Prometheus\footnote{https://prometheus.io/, last accessed: March 22, 2026}), but their attribution heuristics are typically linear and do not model the non-linear power behavior in many utilization regimes.

\textbf{Process- and container-level estimation on single nodes.}
SmartWatts~\cite{fieni_smartwatts_2020} estimates power consumption of processes/containers using calibrated models across CPU frequencies and can improve accuracy compared to purely linear CPU-time models.
In a different but related direction, CodeCarbon~\cite{lacoste_quantifying_2019} and Carbontracker~\cite{anthony_carbontracker_2020} aim to estimate energy and carbon emissions of software runs, often focusing on developer usability and carbon intensity rather than attributing energy to fine-grained tasks in distributed cluster runs.
Such tools are useful for reporting and awareness, but generally do not solve the workflow-task attribution problem.

\textbf{Fine-grained attribution in multi-tenant settings.}
EnergAt~\cite{he_energat_2024} proposes a thread-level, NUMA-aware method for CPU and DRAM attribution that introduces non-linear energy credits. It enables energy attribution for individual processes running on single nodes.
% and not in distributed environments like Kubernetes clusters.
METRION~\cite{weigell_metrion_2025} further studies fine-grained attribution and monitoring for shared environments.
While these approaches address important systems challenges (multi-socket, NUMA, concurrent workloads), they do not target a containerized deployment model and distributed execution in compute clusters, and do not provide workflow-task aggregation for engines such as Nextflow.

\textbf{Positioning of Nf-PEAK.}
Nf-PEAK complements prior work by combining (i) a containerized deployment on Kubernetes, (ii) process-level monitoring using a non-linear credit model, and (iii) task-level aggregation for Nextflow workflows.
This enables practical energy attribution for distributed workflow executions in managed cluster environments where users cannot install node-level tooling.

%\section{Future Work}\label{Future_Work}
%Future work includes reducing the blind spot for sub-second tasks (e.g., through event-based task detection using kernel-assisted mechanisms such as eBPF where permitted), integrating Nf-PEAK more tightly into workflow engines such as Nextflow, and extending attribution to additional components (e.g., accelerators or storage) when reliable counters are available.

\section{Conclusion}\label{Conclusion}
We presented Nf-PEAK, a containerized method for process- and task-level attribution of RAPL-visible CPU and DRAM energy for Nextflow workflows executed on Kubernetes clusters.
By combining RAPL energy counters with per-process performance metrics and a non-linear credit model, Nf-PEAK enables fine-grained energy feedback in managed and multi-tenant environments.
Across three nf-core workflows, Nf-PEAK achieves an average Mean Absolute Percentage Error of 6.6\% in isolated runs and 10.9\% under co-located CPU load on 4 cluster nodes, outperforming Kepler.

In the future, we plan to go beyond measurements and attribution by predicting task-level energy consumption for workflows to enable fine-grained energy and carbon-aware scheduling.

\section*{Data Availability}
Code and usage instructions are available on GitHub at \url{https://github.com/Nf-PEAK/Nf-PEAK_Artifacts}.
%Code and usage instructions are available at \url{https://github.com/CRC-FONDA/Energy-Attribution}.

\section*{Acknowledgments}
Funded by the Deutsche Forschungsgemeinschaft (DFG, German Research Foundation) -- Project-ID 414984028 -- SFB 1404 FONDA.

%\clearpage
\bibliographystyle{plain}
\bibliography{Work}

\end{document}